\documentclass[pra,floatfix,twocolumn]{revtex4}
\usepackage[T1]{fontenc}
\usepackage{amssymb,amsmath}
\usepackage{txfonts}
\usepackage{color,graphicx}

\begin{document}

\title{Potential energy surfaces for electron dynamics
from a model of localized Gaussian wave packets with
valence-bond spin-coupling:
high-harmonic generation spectra
from H and He atoms\footnote{Chem. Phys. \textbf{570}, 111883 (2023).\\
\hspace*{0.4em}https://doi.org/10.1016/j.chemphys.2023.111883}}

\author{Koji Ando\footnote{E-mail: ando\_k@lab.twcu.ac.jp}}

\affiliation{Department of Information and Sciences, 
    Tokyo Woman's Christian University, 
    2-6-1 Zenpukuji, Suginami-ku, Tokyo 167-8585, Japan}

\date{\today}

\begin{abstract}
Potential energy surfaces of electron dynamics (ePES) are constructed from a model of localized electron wave packets (eWP) with non-orthogonal valence-bond (VB) spin coupling and applied to quantum dynamics simulations of high harmonic generation (HHG) spectra of hydrogen and helium atoms induced by intense laser pulses.
The dynamics of the single electron on the ePES is calculated by numerically solving the time-dependent Schr{\"o}dinger equation.
The results reasonably reproduce previous studies.
The dynamics of the electron wave function, dipole moment and dipole acceleration were analyzed by comparing one- and two-dimensional calculations.
It was found that the main part of the wave function remains within a few Bohrs of the nuclear position, while the part of the wave function that is several orders of magnitude smaller in probability density, which escapes the laser-induced potential barrier by quantum tunneling effect, mainly contributes to the HHG.
\end{abstract}

\maketitle

\section{Introduction}
Chemical physics research is now exploring the dynamics of electrons in molecules and materials
\cite{Krause1992,Schafer1993,Corkum1993,Baggesen2011,Bandrauk2013,
Itatani2004,Haessler2011,Salieres2012,Offenbacher2015,Ghimire2010,Yoshikawa2017,
Bucksbaum2007,Kling2008,Krausz2009,Vampa2014,Ramasesha2016,Takatsuka2021}.
In particular, remarkable progress has recently been made in attosecond time-resolved laser spectroscopy.
The high-harmonic generation (HHG) spectra induced by intense laser pulses
are expected to be informative about electron dynamics.
Theoretical simulations of HHG spectra cannot be straightforward extensions of the
conventional time-dependent quantum chemical methods
because of the large spatial domain involved in the electron dynamics
\cite{Grossmann2013,Geppert2008,Takemoto2011,Lostedt2012,
Telnov2013,Tolstikhin2013,Li2016,
Sato2013,Sato2015,Remacle2007,Nest2008,Remacle2011,Nikodem2017,Balzer2010,Ulusoy2012}.
To facilitate large-scale quantum dynamics simulations,
the single active electron (SAE) approximation has often been used.
However, the determination of the functional form and parameters of the single electron effective potential is arbitrary.
Research has been carried out to derive accurate effective potentials for single electron dynamics,
which would provide the theoretical basis for the SAE model
\cite{Tong2005,Ohmura2017,Kato2018,Kocak2020,Schild2017}.

As an approach to this problem, we study a model of floating and breathing localized electron wave packets (eWP) with valence-bond (VB) spin coupling
\cite{Ando2009,Ando2012,Ando2016,Ando2017,Ando2018,Ando2020}.
The model correctly describes the electronic ground states of H$_2$, LiH, CH$_2$, NH$_3$ and H$_2$O
\cite{Ando2012}.
The electronic excited state energies obtained by quantizing the electron potential energy surfaces (ePES) as a function of the eWP center positions were qualitatively appropriate for the singlet and triplet $\Sigma$ and $\Pi$ states of LiH
\cite{Ando2016}.
Moreover, the HHG spectrum calculated from the single-electron dynamics on the ePES of LiH reproduced the more elaborate time-dependent complete-active-space calculation
\cite{Ando2017,Ando2018,Ando2020}.

In this paper we study the HHG spectra of hydrogen and helium atoms.
The adequacy of the ePES model for these simple atoms is not trivial, since the interaction between the active electron and the nuclear charge is not shielded by the core electrons.
In particular, the two electrons in the helium atom are equivalent in the unperturbed state of the (1s)$^2$ configuration, but the symmetry would be broken when responding to the strong external field.
It is therefore not obvious whether the ePES model can describe the electron dynamics under the intense laser field for HHG.

We note that the present HHG calculations are on single atoms with the dipole approximation.
More realistic picture should be obtained from the electromagnetic field propagation 
coupled with the macroscopic polarization of many atoms in the gaseous media. 
This would be considered by the classical Maxwell equations
or by the quantum electrodynamics \cite{Gorlach2020}. 
This subject is interesting and important, but is beyond the scope of this paper.

In Sec. \ref{sec:model} the theoretical model and the calculation method are outlined.
The results of the calculations are discussed in Sec. \ref{sec:results}.
Section \ref{sec:concl} concludes.

\section{Model and Computation}
\label{sec:model}
The theory and computation are basically the same as those in our previous publications on LiH molecule \cite{Ando2018,Ando2020}.
In the VB eWP model, the total electronic wave function is represented by an antisymmetrized product of spatial and spin functions,
\begin{equation}
\psi(1,\cdots,N) = {\cal A}[\phi_1(\boldsymbol{r}_1)\cdots\phi_N(\boldsymbol{r}_N)\theta(1,2)\cdots\theta(N-1,N)] ,
\label{eq:VBwfn}
\end{equation}
in which $\boldsymbol{r}_i$ is the spatial coordinate of $i$th electron.
The one electron spatial function $\phi_i (\boldsymbol{r}_i)$ is a spherical Gaussian WP 
with variable central position $\boldsymbol{q}_i$ and width $\rho_i$,
\begin{equation}
\phi_i(\boldsymbol{r}) = (2\pi\rho_i^2)^{-3/4}\exp(-|\boldsymbol{r}-\boldsymbol{q}_i|^2/4\rho_i^2) ,
\end{equation}
The spin part is the perfect-pairing form with $\theta(i,j)=(\alpha(i)\beta(j)-\beta(i)\alpha(j))/\sqrt{2}$.

The electronic energy, $E=\langle \psi|\hat{H}|\psi\rangle/\langle \psi | \psi \rangle$, 
is minimized as
a function of the variables 
$\boldsymbol{q}_1, \cdots, \boldsymbol{q}_N$ and $\rho_1,\cdots, \rho_N$, 
to determine the optimal values 
$\boldsymbol{q}_1^{(0)}, \cdots, \boldsymbol{q}_N^{(0)}$ and $\rho_1^{(0)},\cdots, \rho_N^{(0)}$
in the ground state.
The effective potential function for the $i$th electron ${\cal V}_i(\boldsymbol{q})$,
which we call the ePES,
is constructed by fixing the variables other than $\boldsymbol{q}_i$ at the optimized values:
\begin{equation}
{\cal V}_i(\boldsymbol{q}) = E(\boldsymbol{q}_1^{(0)}, \cdots, 
\boldsymbol{q}_{i-1}^{(0)}, \boldsymbol{q}, \boldsymbol{q}_{i+1}^{(0)}, 
\cdots, \boldsymbol{q}_N^{(0)},
\rho_1^{(0)},\cdots, \rho_N^{(0)})
\end{equation}
The time-dependent Schr{\"o}dinger equation for a single electron on the ePES 
is numerically solved to compute the HHG spectrum.

\subsection{Hydrogen-like atoms}
For hydrogen-like atoms with the potential energy
\begin{equation}
V(\boldsymbol{r}) = -Z/|\boldsymbol{r}| ,
\end{equation}
where $Z$ is the nuclear charge,
the ePES is given analytically by \cite{Ando2009}
\begin{equation}
{\cal V}(\boldsymbol{q})
= -\sqrt{\frac{2}{\pi}}\frac{Z}{\rho^{(0)}} F_0\left(\frac{|\boldsymbol{q}|^2}{2(\rho^{(0)})^2}\right)
\label{eq:ePEShatom}
\end{equation}
where 
$F_0$ is the Boys function of order 0
\begin{equation}
F_0(t)
= \sqrt{\frac{\pi}{4t}}\mathrm{erf}(\sqrt{t}) .
\end{equation}
The optimal eWP width is given by
\begin{equation}
\rho^{(0)} = \frac{3}{4Z}\sqrt{\frac{\pi}{2}}
\end{equation}
(Note that there is a typo in Eq. (10) of Ref. \cite{Ando2009}.)
$\rho^{(0)} = 0.940$ bohr for $Z=1$ (H atom) and 0.470 bohr for $Z=2$ (He$^{+}$ ion).
The optimal eWP center is located at the nuclear position, $\boldsymbol{q}^{(0)} = 0$.

\subsection{Helium atom}
For helium atom, Eq. (\ref{eq:VBwfn}) has the Heitler-London form,
\begin{equation}
\psi(1,2) = N
(\phi_1(\boldsymbol{r}_1)\phi_2(\boldsymbol{r}_2) + \phi_2(\boldsymbol{r}_1)\phi_1(\boldsymbol{r}_2))
(\alpha(1)\beta(2)-\beta(1)\alpha(2)) ,
\label{eq:HL}
\end{equation}
where $N$ is the normalization factor.
The numerical energy minimization gives
the electronic energy $E=-2.557$ hartrees
with 
$\rho_1^{(0)} = 0.833$ bohr and $\rho_2^{(0)} = 0.397$ bohr, and
the eWP centers located at the nuclear position, $\boldsymbol{q}_1^{(0)} = \boldsymbol{q}_2^{(0)} = 0$.
As the eWP model gives $-1.698$ hartrees for the energy of He$^{+}$ ion (Eq. (11) of Ref. \cite{Ando2009}),
the ionization potential (IP) is 0.859 hartree.
These are comparable to the reference values,
$E(\mathrm{He}) = -2.903$ hartrees,
$E(\mathrm{He}^{+}) = -2.000$ hartrees,
and IP = 0.903 hartree \cite{Kandula2010}.
Notably, the VB function consists of two eWP of different widths,
$\rho_1^{(0)}$ and $\rho_2^{(0)}$. 
As will be discussed in Sec.\ref{sec:resultsHe}, 
the ePES for the larger eWP has the smaller binding energy, 
and appropriately represents the SAE dynamics under the strong laser field.

\subsection{Electron dynamics and HHG}
The electron dynamics are induced by a laser pulse with time-dependent electric field of 
\begin{equation}
{\cal E}(t) = {\cal E}_0 \sin(\omega_0 t) \sin^2(\pi t/\tau) \quad (0 \le t \le \tau) ,
\label{eq:laserfield}
\end{equation}
parallel to the $x$-axis.
The frequency $\omega_0$, the field intensity ${\cal E}_0$, and the pulse duration time $\tau$ 
are the parameters to be specified in Sec. \ref{sec:results}.
The atom was placed at the origin, and
the edges of the simulation box were at $\pm 400$ bohrs for both one- and two-dimensional calculations.
The transmission-free absorbing potential of 80 bohrs length was set at each edge.
The initial wave function at $t=0$ was the numerical solution of the time-independent
Sch{\"o}dinger equation for the soft-core (Eq.(\ref{eq:vsoft}))
and the cusp (Eq.(\ref{eq:vcusp})) models,
and the Gaussian WP with the optimized parameters
$\boldsymbol{q}$ and $\rho$
for the ePES model.
The wave function was propagated with Cayley's hybrid scheme \cite{Watanabe00}
with the spatial grid length of 0.2 bohr
(4000 points per direction)
and the time-step of 0.01 a.u. ($\simeq$ 0.24 as).
The HHG spectra were computed from the Fourier transform 
with the Hanning window of the dipole acceleration dynamics.

Although the eWP 
for constructing the ePES 
are spherical in three dimension, 
the quantum dynamics calculations in this work were limited to one and two dimensions,
mainly due to the computational cost associated with the large box length required for 
describing the dynamics induced by the strong laser pulse.

\section{Results and Discussion}
\label{sec:results}
\subsection{Hydrogen atom}
\subsubsection{Electron potential under laser field}

\begin{figure}[h]
\centering
\includegraphics[width=0.35\textwidth]{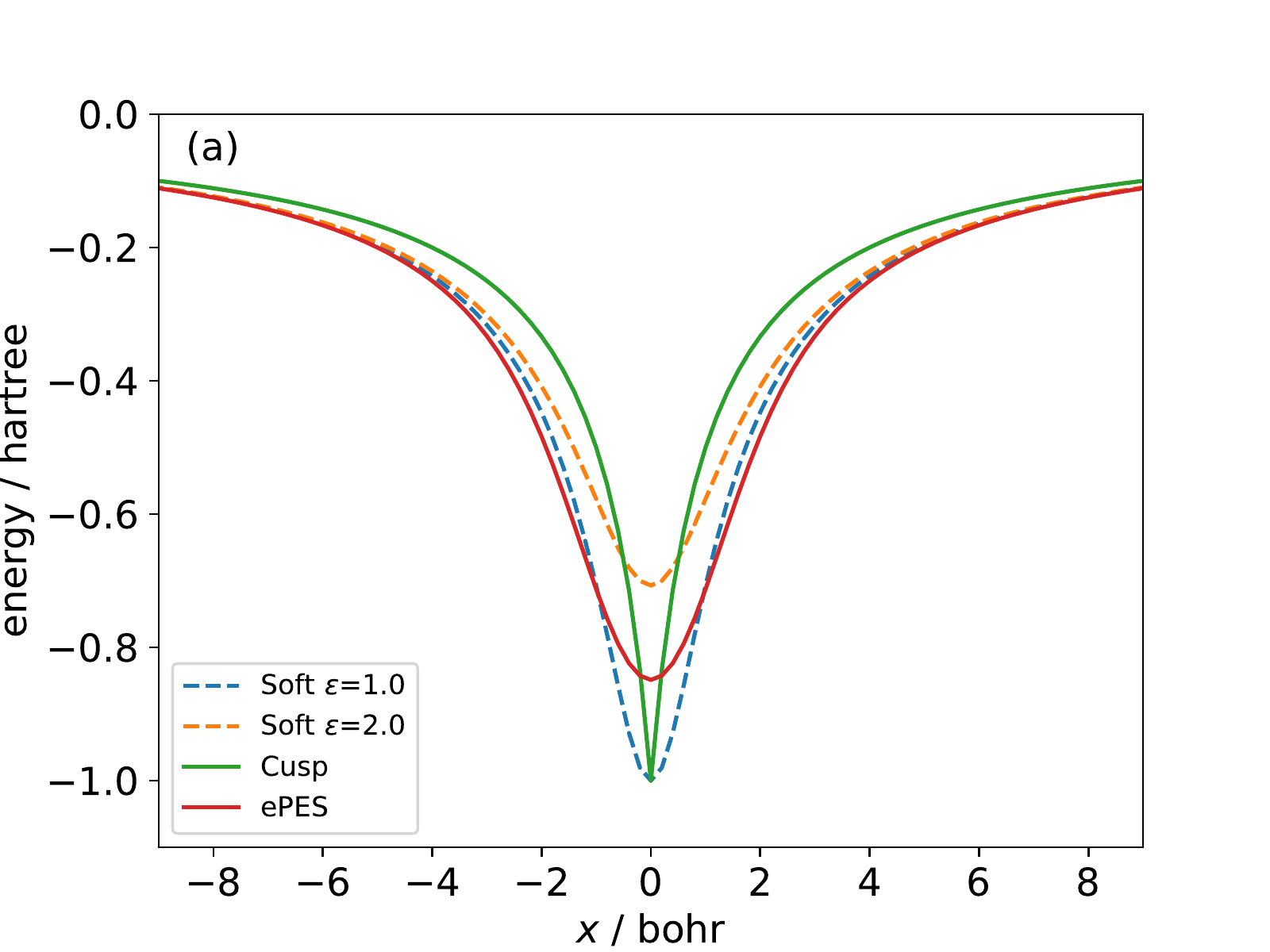}
\includegraphics[width=0.35\textwidth]{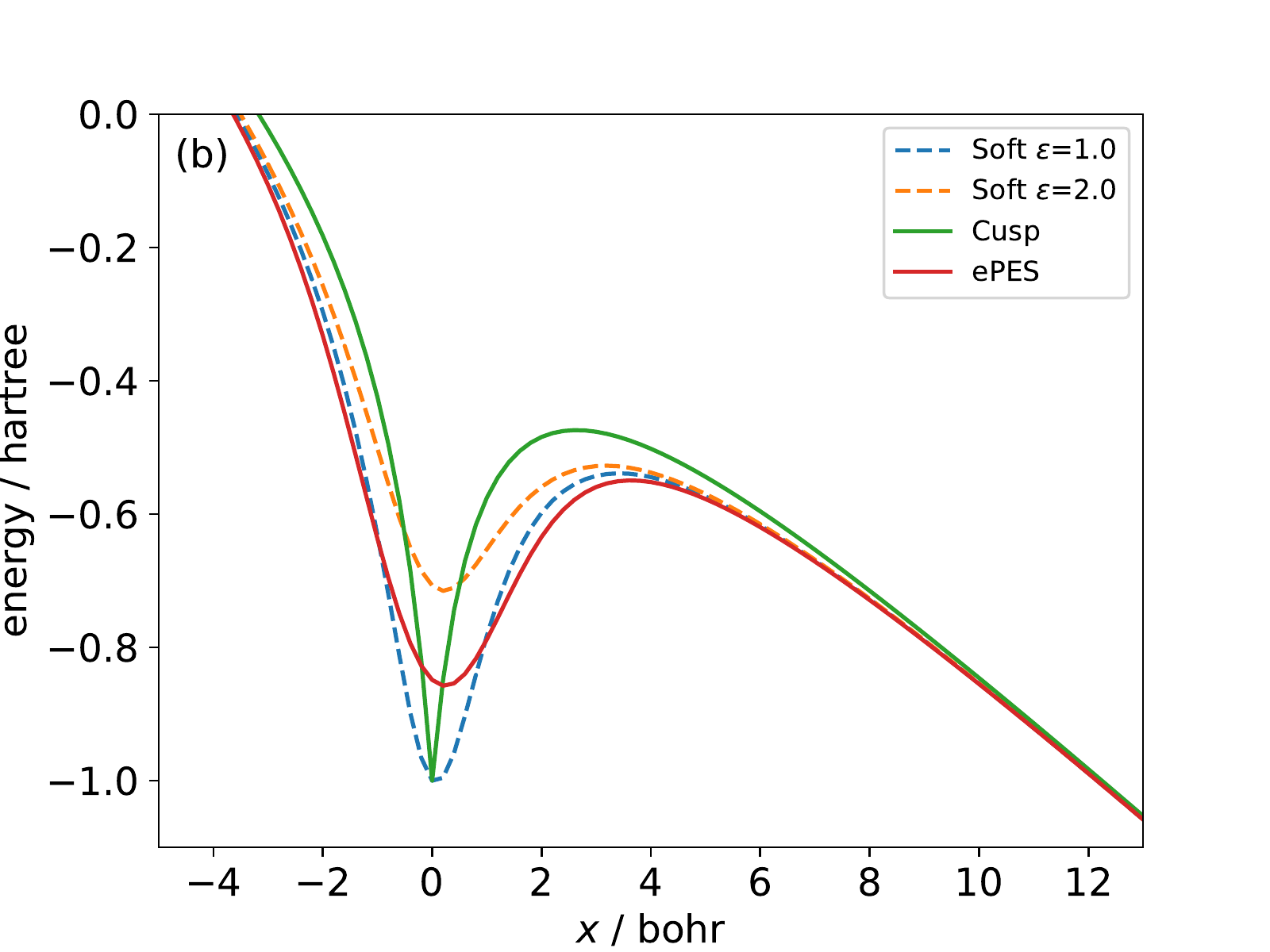}
\caption{Potential energy curves for an electron in hydrogen atom;
    the soft-core model Eq. (\ref{eq:vsoft}) with the parameter $\epsilon=$ 1.0 and 2.0, 
    the cusp model Eq. (\ref{eq:vcusp}), 
    and the ePES, (a) without and (b) under the maximum laser field.}
\label{fig:h1dV}
\end{figure}

\begin{figure}[h]
\centering
\includegraphics[width=0.35\textwidth]{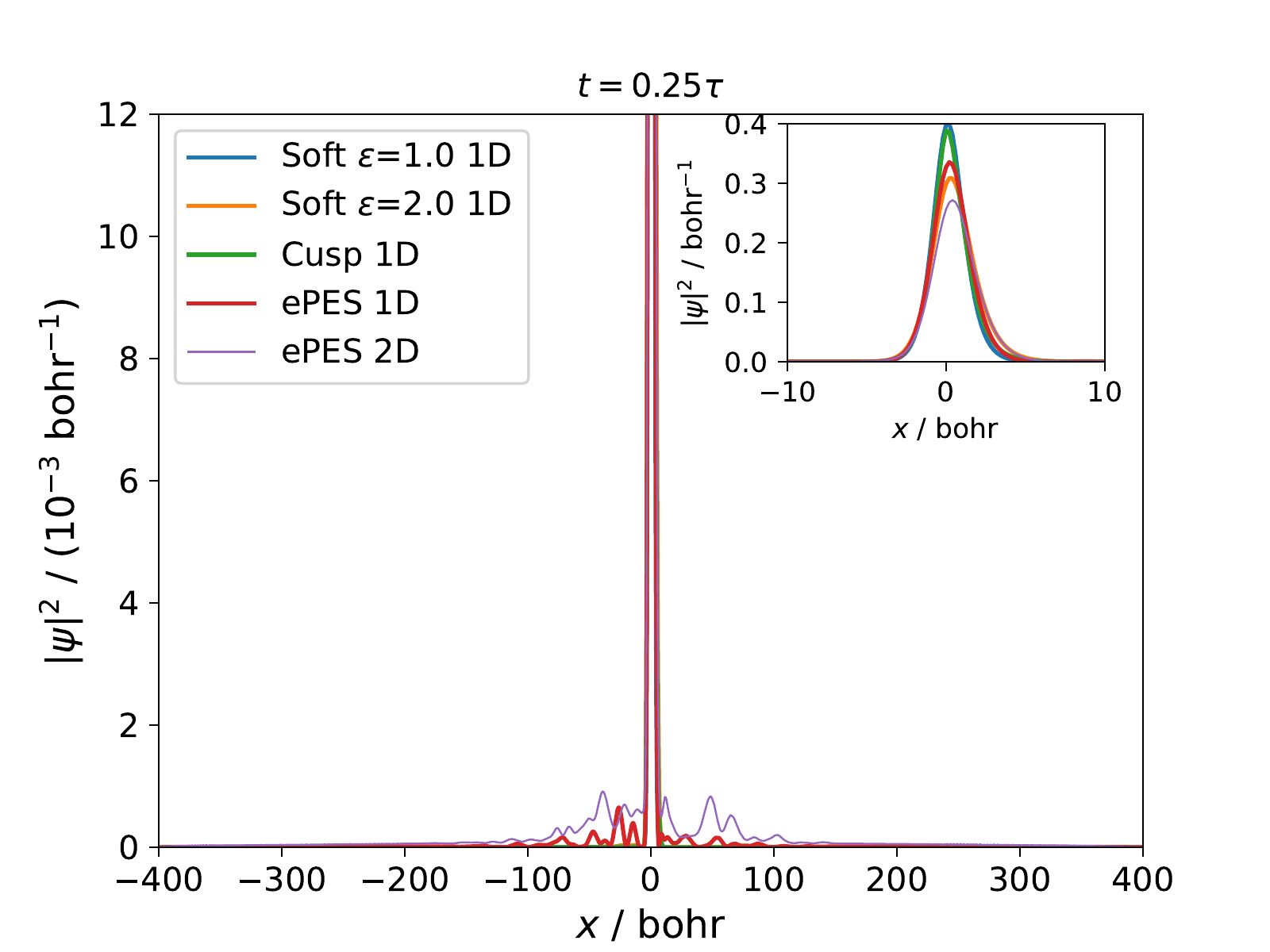}
\includegraphics[width=0.35\textwidth]{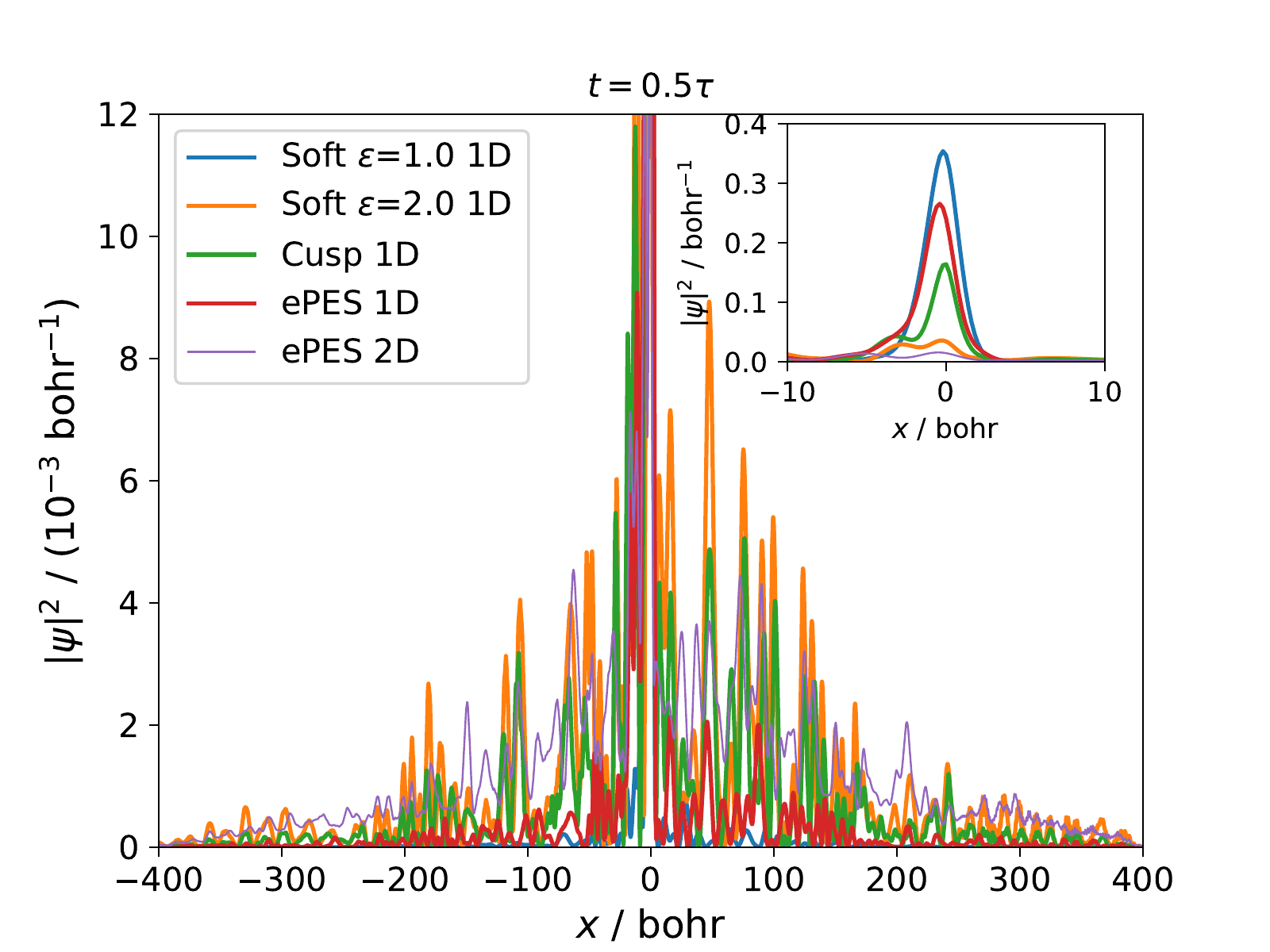}
\includegraphics[width=0.35\textwidth]{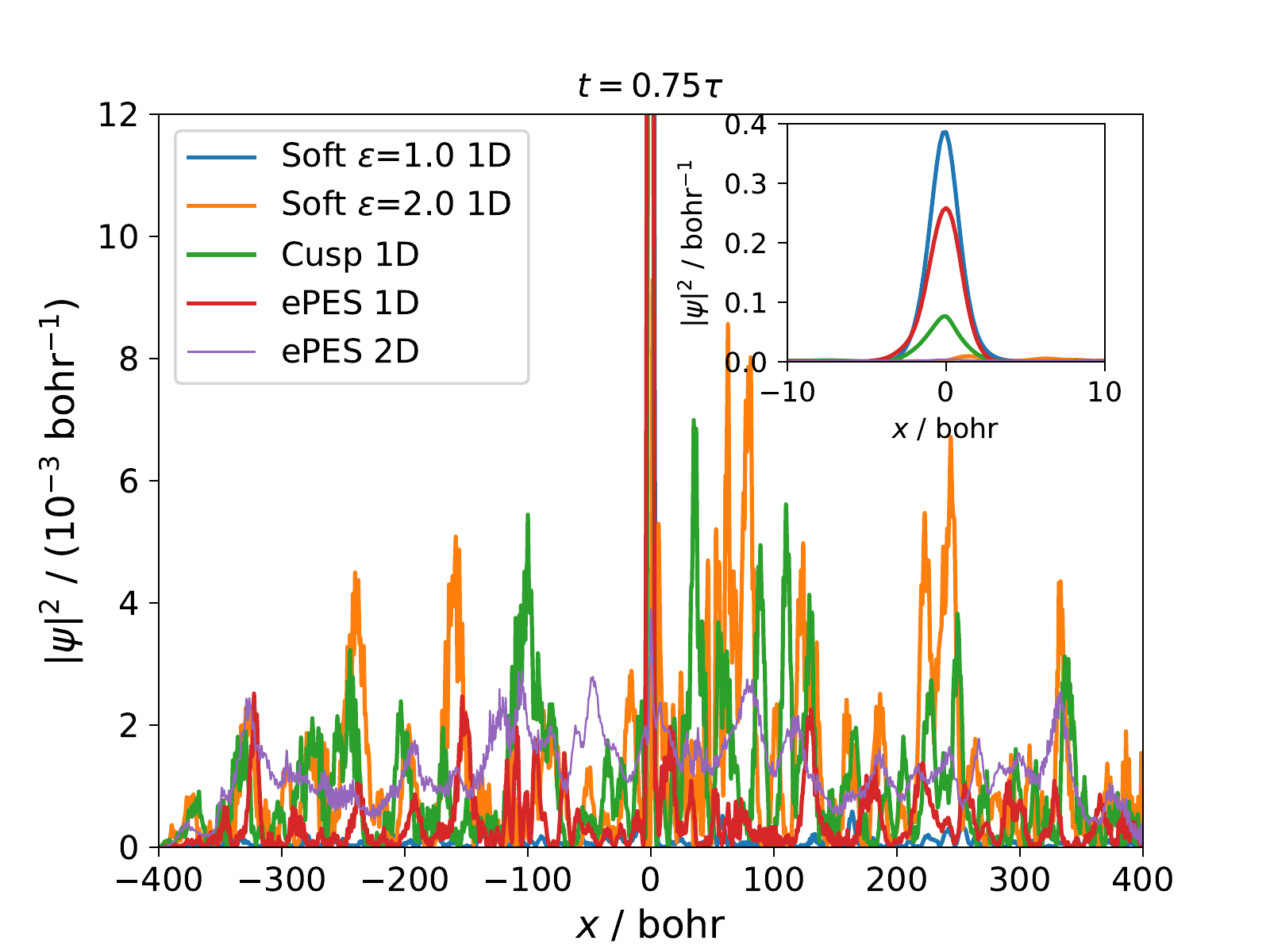}
\caption{
Electron probability density of hydrogen atom at $t=0.25\tau, 0.5\tau$ and 0.75$\tau$
from one-dimensional calculations
with the soft-core potential Eq. (\ref{eq:vsoft}) with $\epsilon=$ 1.0 and 2.0,
the cusp potential Eq. (\ref{eq:vcusp}), and the ePES.
The probability density of two-dimensional ePES model was integrated along the $y$ direction.}
\label{fig:hWfn}
\end{figure}

Figure \ref{fig:h1dV} (a) compares the ePES with commonly used potential models. 
The soft-core potential \cite{Rae1994,Bauer1997,Lein2000,Hu2004} has a form
\begin{equation}
V_\mathrm{soft}(x) = -\frac{Z}{\sqrt{x^2+\epsilon}}
\label{eq:vsoft}
\end{equation}
in which $\epsilon$ is the parameter to remove the singularity of the Coulomb potential.
The cusp model defined by
\begin{equation}
V_\mathrm{cusp}(x) = -\frac{Z}{1+|x|}
\label{eq:vcusp}
\end{equation}
has been demonstrated to be appropriate for computation of HHG \cite{Gordon2005PRA}.
As seen in Fig. \ref{fig:h1dV},
the ePES model does not have a cusp, and 
the depth is intermediate between the soft-core potentials with $\epsilon=$ 1.0 and $\epsilon=$ 2.0.
The cusp potential is as deep as the soft-core potential with $\epsilon=$ 1.0, but is notably narrower.
The shapes of the potentials will be related to the tunneling distribution in Fig. \ref{fig:hWfn}
and the dynamics of the dipole and the dipole acceleration in Fig. \ref{fig:h1d2dDplt}.

Figure \ref{fig:h1dV} (b) displays the potential energy curves under the maximum laser field ${\cal E} = {\cal E}_0$.
At ${\cal E}=-{\cal E}_0$, the curve $V(x) + x {\cal E}_0$ is the symmetric image of Fig. \ref{fig:h1dV} (b) about $x=0$.
The wave function tunnels out through the energy barrier in both direction.
The height and width of the barrier are in ranges 0.2--0.6 hartree and 8--12 bohr.

The laser pulse parameters in Eq. (\ref{eq:laserfield}) 
for the calculation of HHG from hydrogen atom is taken from Ref. \cite{Hu2004}.
The frequency $\omega_0$ corresponding to the wavelength 800 nm, 
the intensity $I_0 = 2 \times 10^{14} \;\mathrm{W/cm^2}$
corresponding to
${\cal E}_0 = 3.882 \times 10^{8} \;\mathrm{V/cm}$
($=7.549 \times 10^{-2}$ atomic units),
and the pulse duration $\tau = 50 \;\mathrm{fs}$ corresponding to 18.7 cycle.
This setting gives the ponderomotive energy of $U_p = 0.439$ hartree 
that corresponds to 7.7 harmonic orders.

\subsubsection{High-harmonic generation spectra}

\begin{figure}[h]
\centering
\includegraphics[width=0.35\textwidth]{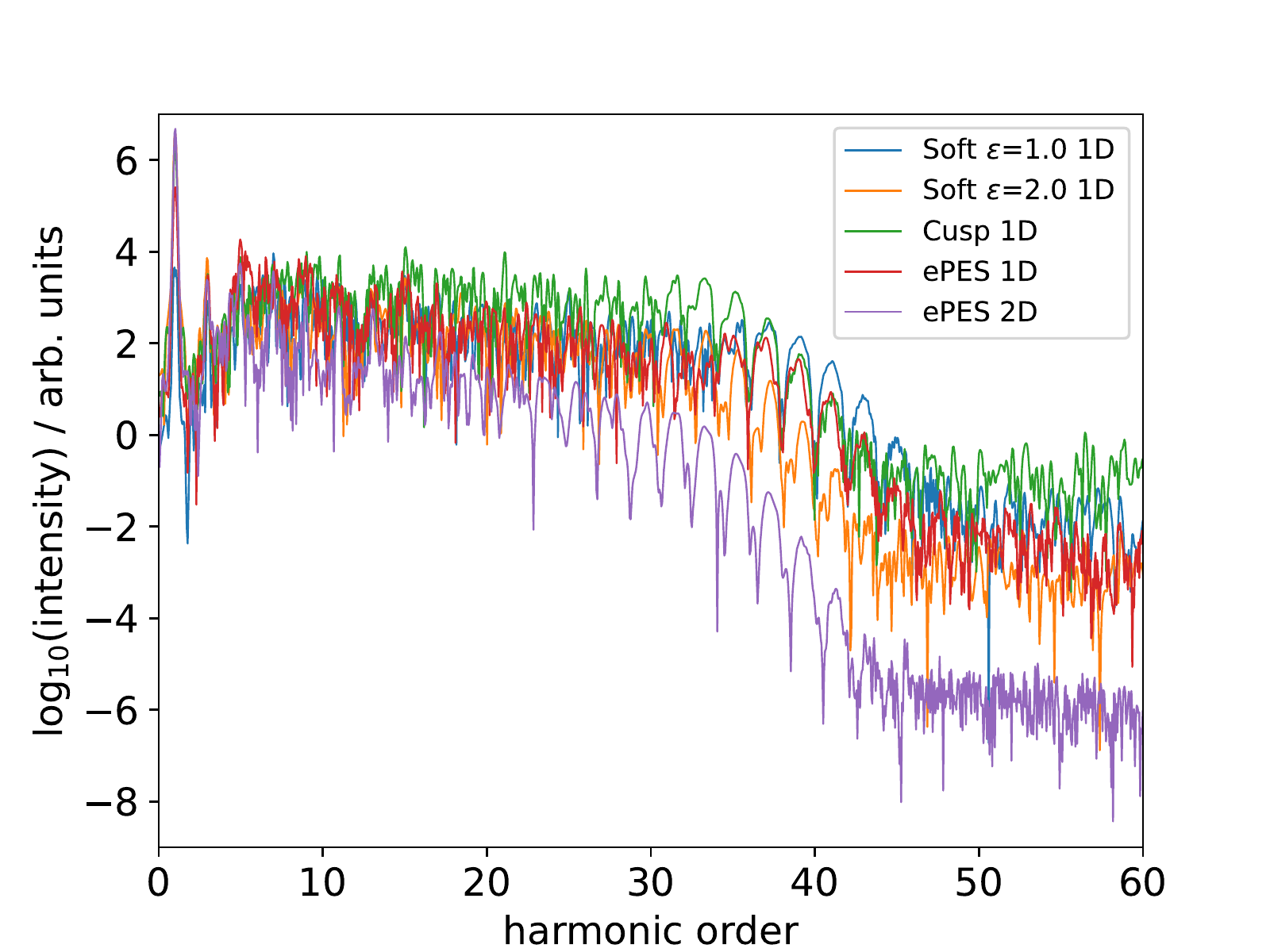}
\caption{High-harmonic generation spectra from hydrogen atom with 
    the one-dimensional soft-core potential Eq. (\ref{eq:vsoft}) with $\epsilon=$ 1.0 and 2.0, 
    the one-dimensional cusp model Eq. (\ref{eq:vcusp}),
    and the one- and two-dimensional ePES.}
\label{fig:h1d2dHHG}
\end{figure}

Figure \ref{fig:h1d2dHHG} displays the computed HHG spectra.
All spectra exhibit a cut-off 
at $\sim$35 harmonic orders corresponding to $\sim$54 eV,
in agreement with the previous works \cite{Hu2004}.
This is also consistent with the estimate by the formula 
$\mathrm{IP} + 3.2 U_p = 34$ harmonic orders,
with the IP of hydrogen atom = 0.5 hartree = 8.8 harmonic orders.
The spectra of the one-dimensional models agree well.
Dependence on the parameter $\epsilon$ of the soft-core potential is not large,
and the existence of potential cusp does not seem critical in the present setting.
The spectrum of the two-dimensional ePES model deviates from the others
in the region higher than $\sim$20 harmonic orders.
This relates to the tunneling probability density
discussed next.

\begin{figure}[h]
\centering
\includegraphics[width=0.35\textwidth]{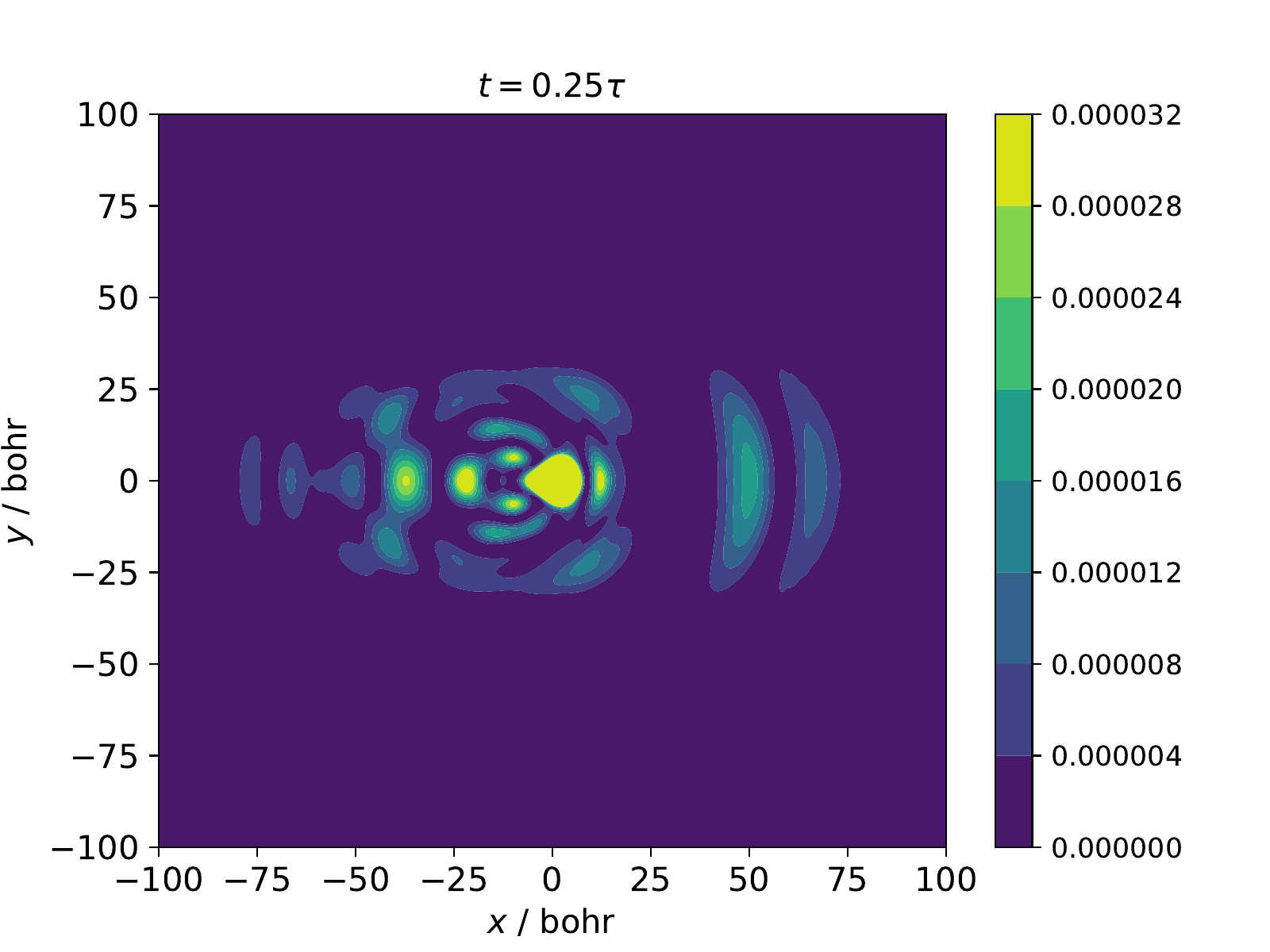}
\includegraphics[width=0.35\textwidth]{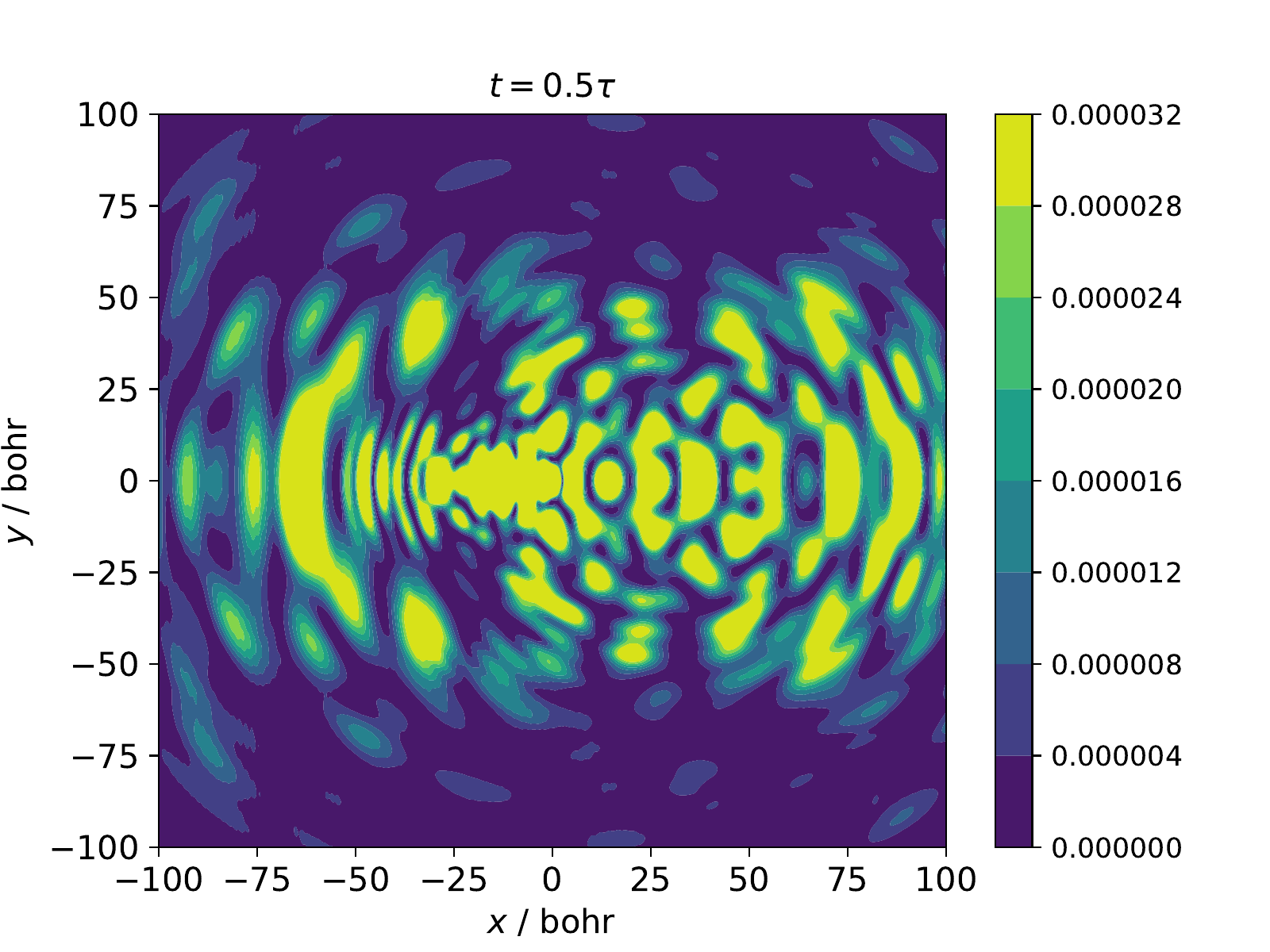}
\includegraphics[width=0.35\textwidth]{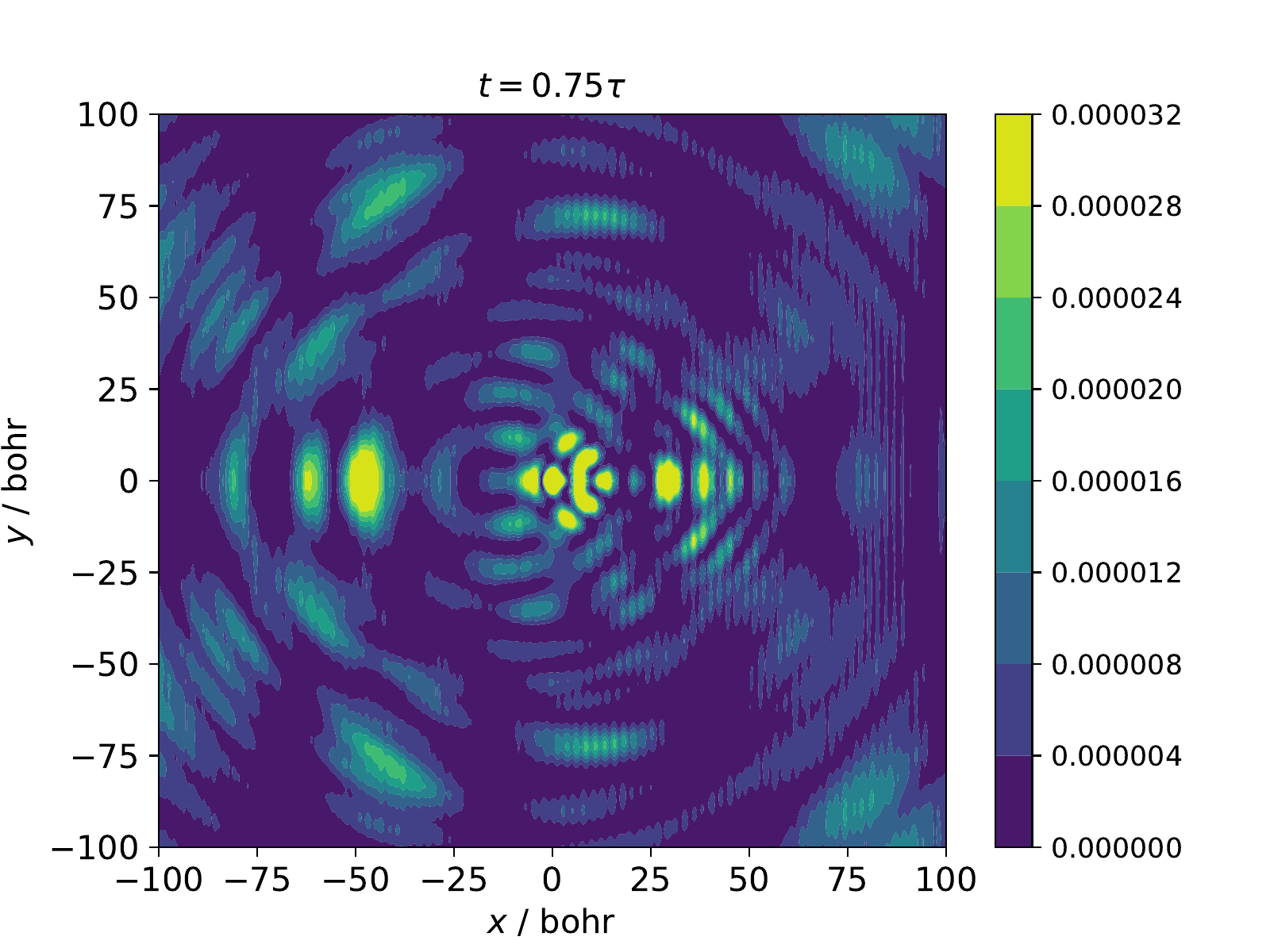}
\caption{
Electron probability density of hydrogen atom 
at $t=0.25\tau$, $0.5\tau$ and $0.75\tau$
for the two-dimensional ePES model,
truncated at a maximum value set to $3\times 10^{-5}$ bohr$^{-2}$
to see the small component escaped from the potential well by the laser-induced quantum tunneling.}
\label{fig:h2dWfn}
\end{figure}

\subsubsection{Tunnel probability density}

Figure \ref{fig:hWfn} displays the electron probability density at $t=0.25\tau$, $0.5\tau$, and $0.75\tau$.
The data from the two-dimensional calculation was integrated in the $y$ direction.
The insets show the whole profiles 
which indicate that the main part of the probability density remains localized within a few bohrs
near the nucleus.
Note the orders of magnitude difference in the scales of the vertical and horizontal axes 
between the main plots and the insets.
The figures indicate that 
the electron probability density approximately two orders of magnitude smaller 
than the main peak
escape through the potential barrier (see Fig. \ref{fig:h1dV} (b)) by laser-induced quantum tunneling 
and spread over several hundred bohrs.

The tunnel density at $t = 0.25\tau$ is notable only for 
the one- and two-dimensional ePES model.
The two-dimensional ePES model always show higher tunnel density 
compared to the one-dimensional ePES model.
This will be related to the two-dimensional plot in Fig. \ref{fig:h2dWfn};
the tunneling would be suppressed by the dimensional restriction.
The smaller tunnel density at $t=0.25\tau$ for the soft-core model and the cusp model
might be related to the tunnel length seen in Fig. \ref{fig:h1dV} (b).
These latter models have tunnel densities in $t \ge 0.5\tau$,
but that for the soft-core with $\epsilon=$ 1.0 is notably smaller.
This might be related to the depths and the widths of the potentials
in Fig. \ref{fig:h1dV}.

Figure \ref{fig:h2dWfn} displays the electron probability density 
at $t=0.25\tau$, $0.5\tau$, and $0.75\tau$
for the two-dimensional ePES model.
The data was
truncated at a maximum value set to $3\times 10^{-5}$ bohr$^{-2}$
to see the small component outside the potential well.
Initially, the nodal structure of the wave function develops mainly in the $x$-direction 
along the laser electric field.
Meanwhile,
the wave function expands in the $y$-direction, 
eventually to an almost isotropic nodal structure.
The component distributed toward $y$-direction
appears to suppress the HHG
compared to the one-dimensional calculations
in the region higher that $\sim$30 harmonic orders as seen in Fig. \ref{fig:h1d2dHHG}

\subsubsection{Dipole dynamics and spectral evolution}

\begin{figure}[h]
\centering
\includegraphics[width=0.35\textwidth]{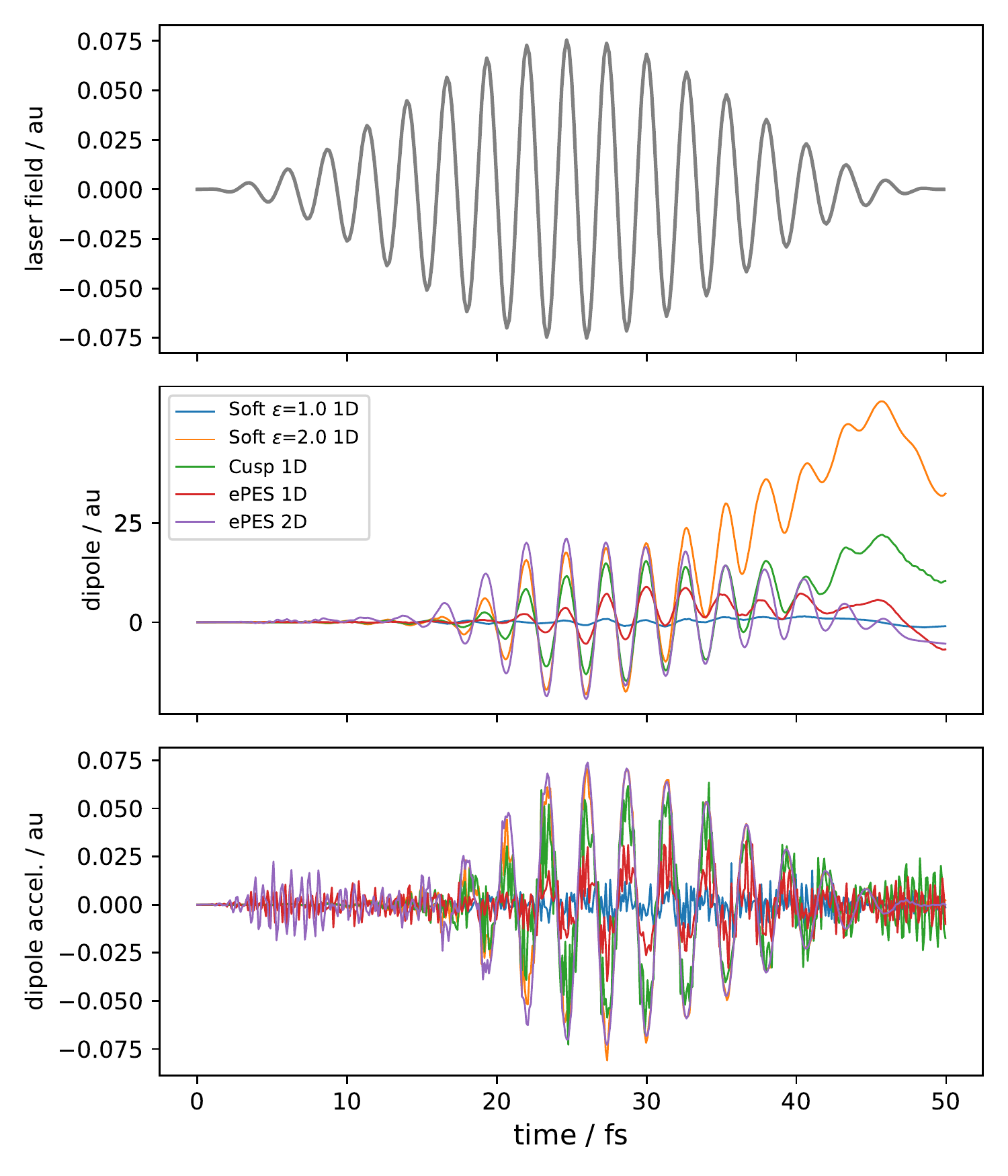}
\caption{
Time profiles of the laser field (top) and the dipole moment (middle)
and the dipole acceleration (bottom) of hydrogen atom,
with the soft-core potential Eq. (\ref{eq:vsoft}) with $\epsilon=$ 1.0 and 2.0, 
the cusp model Eq. (\ref{eq:vcusp}),
and the one- and two-dimensional ePES.}
\label{fig:h1d2dDplt}
\end{figure}

Figure \ref{fig:h1d2dDplt} displays the time profiles of the laser pulse, 
the dipole moment, and the dipole acceleration.
The oscillation amplitude of the dipole and the dipole acceleration are largest for 
the two-dimensional ePES model, followed by the soft-core model with $\epsilon=$ 2.0, the cusp model,
the one-dimensional ePES, and the soft-core model with $\epsilon=$ 1.0.
This order appears be related to the amplitude of the tunneling distribution
discussed in Fig. \ref{fig:hWfn}.
The dipole moment of the soft-core model with $\epsilon=$ 2.0 starts to diverge after $t \simeq 30$ fs.
That of the cusp model also deviate after $t \simeq$ 40 fs.
These are also related to the shapes of the potential curves
in Fig. \ref{fig:h1dV}.

\begin{figure}[h]
\centering
\includegraphics[width=0.35\textwidth]{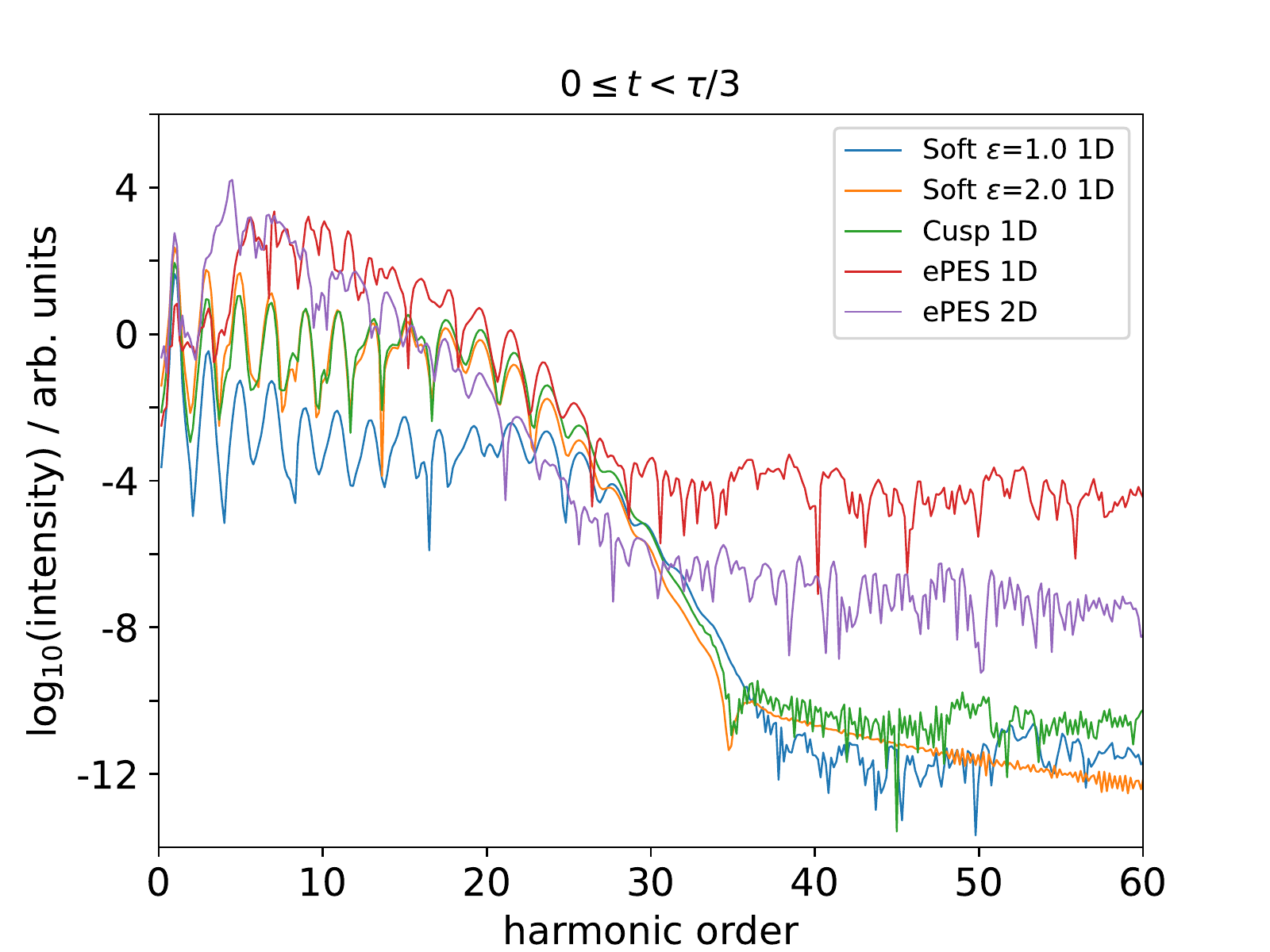}
\includegraphics[width=0.35\textwidth]{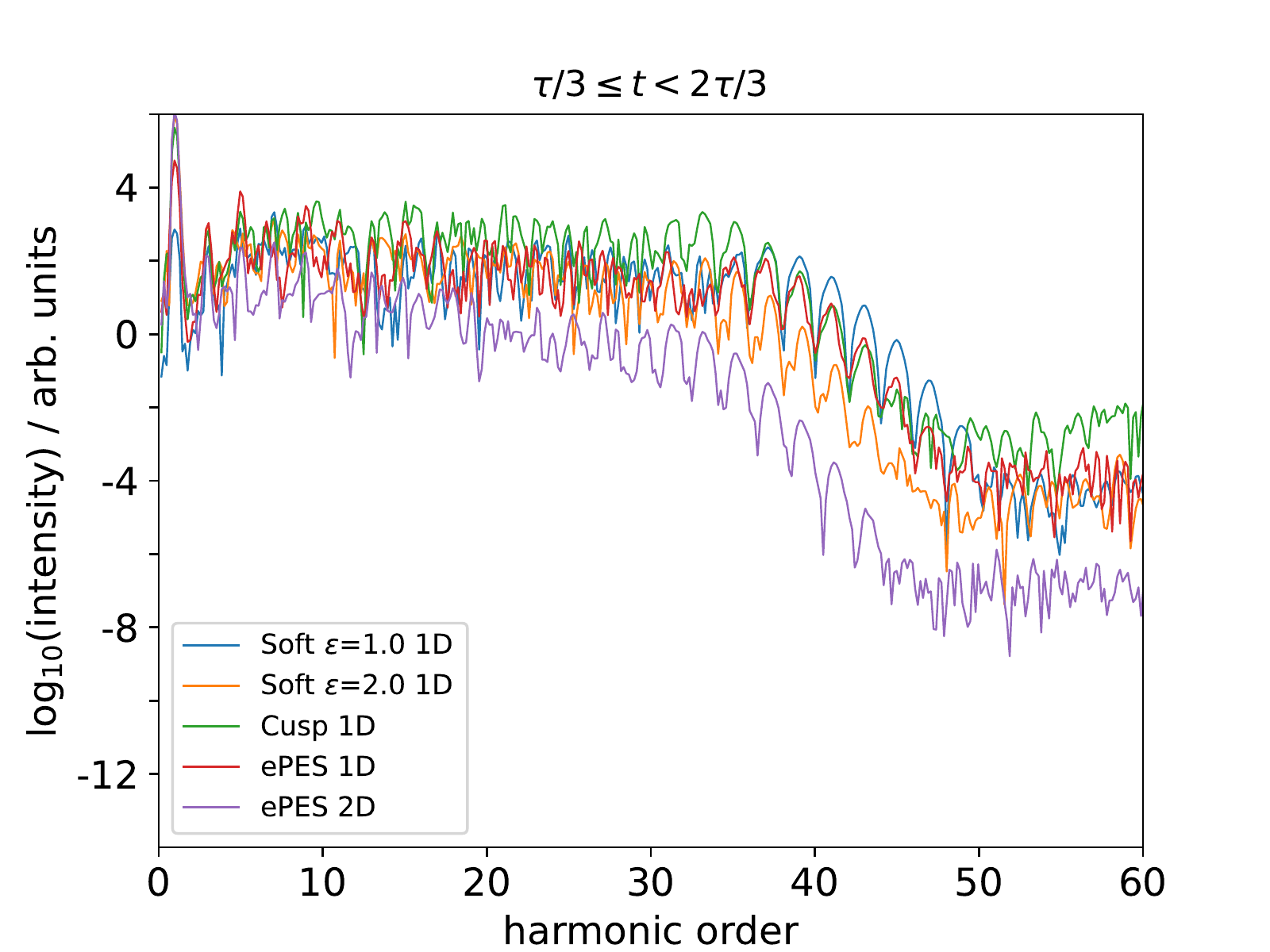}
\includegraphics[width=0.35\textwidth]{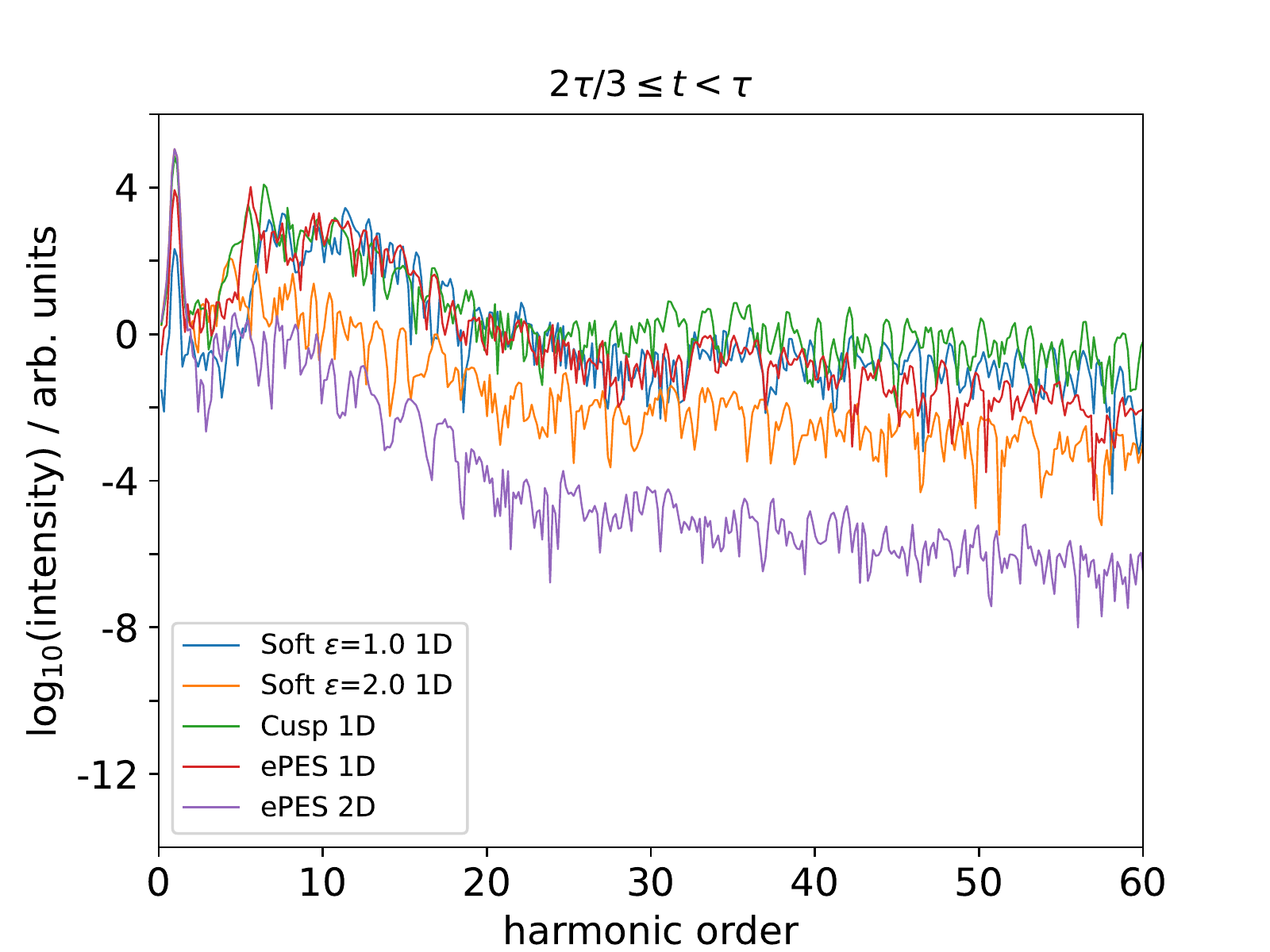}
\caption{High-harmonic generation spectra of hydrogen atom,
similar to Fig. \ref{fig:h1d2dHHG} but 
computed from three time windows,
$0 \le t < \tau/3$, $\tau/3 \le t < 2\tau/3$, and $2\tau/3 \le t < \tau$.}
\label{fig:h1dHHG3tw}
\end{figure}

To gain further insight, 
we divided 
the dipole acceleration trajectories into three time windows,
$0 \le t < \tau/3$, $\tau/3 \le t < 2\tau/3$, and $2\tau/3 \le t < \tau$,
and computed the HHG spectra.
The results are displayed in Fig. \ref{fig:h1dHHG3tw}.
The difference among the models is most apparent in the spectra from the
shortest time window, $0 \le t < \tau/3$,
in which the one- and two-dimensional ePES models have larger intensities
compared to the other models.
This would be related to the tunnel density at $t = 0.25 \tau$ seen in Fig. \ref{fig:hWfn}.
In $t \ge \tau/3$, the intensity for the two-dimensional ePES model is the smallest,
presumably for the same reason as discussed on Fig. \ref{fig:h1d2dHHG}.
The smaller intensity for the soft-core model with $\epsilon=2.0$
in $2\tau/3 \le t < \tau$ would be related to the behavior of the dipole trajectory
seen in Fig. \ref{fig:h1d2dDplt}.

\subsection{Helium atom}
\label{sec:resultsHe}
\subsubsection{Electron potential under laser field}

\begin{figure}[h]
\centering
\includegraphics[width=0.35\textwidth]{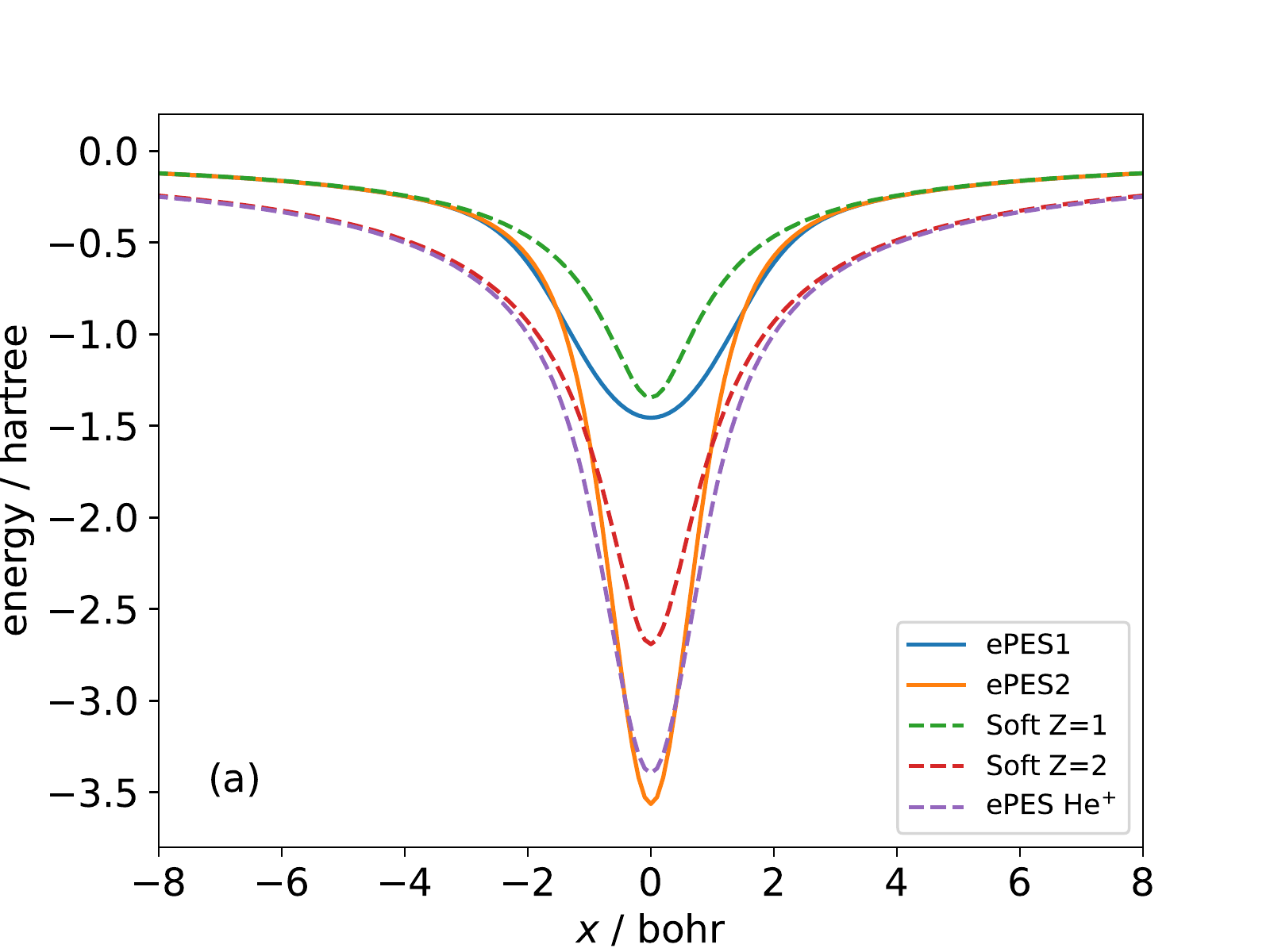}
\includegraphics[width=0.35\textwidth]{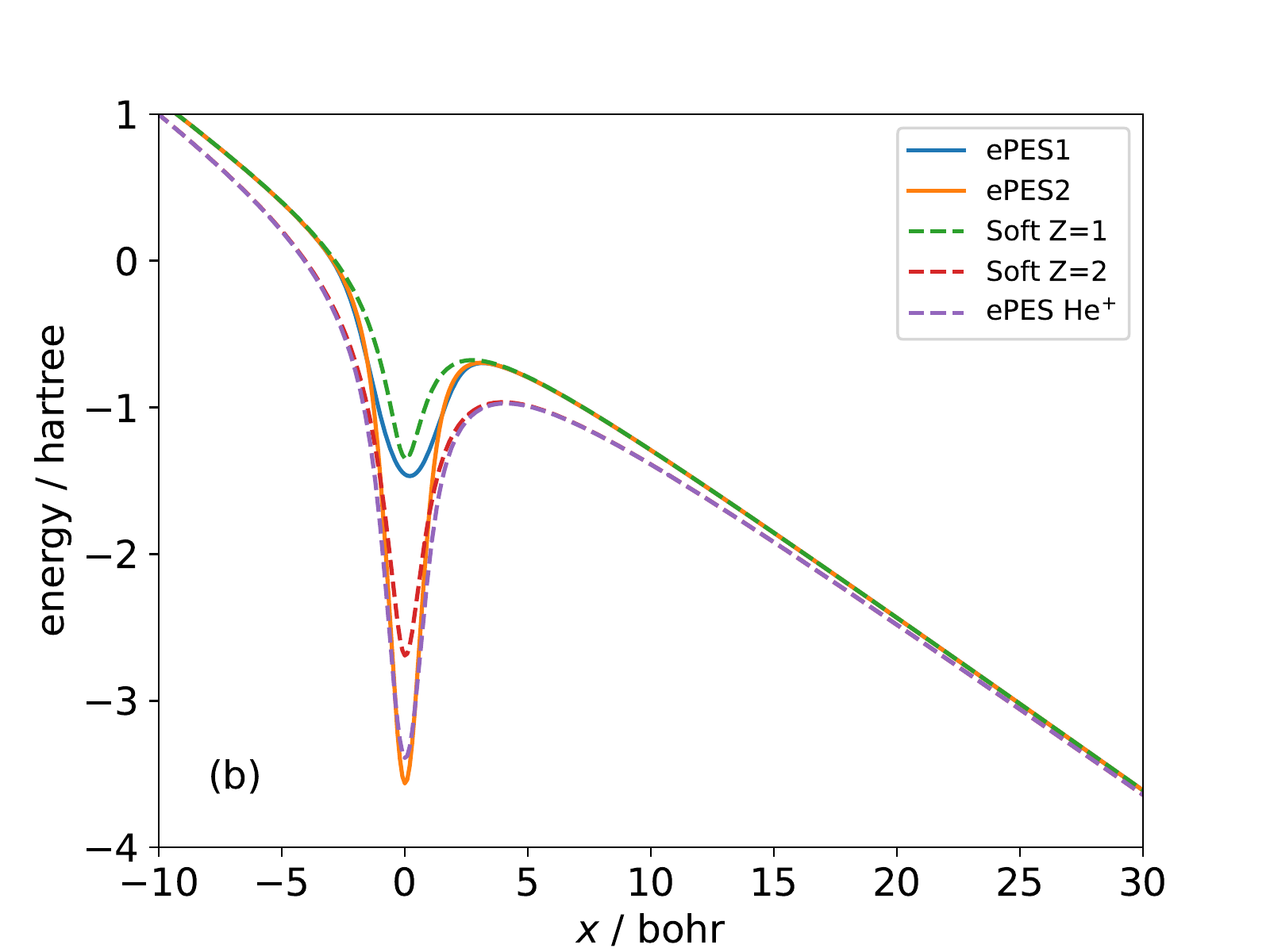}
\caption{Potential energy curves for electrons in helium atom;
    ePES1 and ePES2 corresponding to the eWP widths
    $\rho_1^{(0)}$ and $\rho_2^{(0)}$, respectively,
    the soft-core potential Eq. (\ref{eq:vsoft}) with $\epsilon=0.55$ 
    and the nuclear charge $Z=$ 1 and 2,
    and the ePES for He$^{+}$ ion, (a) without and (b) under the maximum laser field.}
\label{fig:he1dV}
\end{figure}

We now present the results for the helium atom.
Figure \ref{fig:he1dV} displays the ePES for the two electrons in helium atom;
ePES1 from $\phi_1$ with $\rho_1^{(0)}=0.833$ bohr
and
ePES2 from $\phi_2$ with $\rho_2^{(0)}=0.397$ bohr.
They are compared with the soft-core potential Eq. (\ref{eq:vsoft}) with $\epsilon = 0.55$,
which was optimized in Ref. \cite{Bauer1997} for a one-dimensional two electron model of helium atom, 
that gives the electronic energy $E = -2.897$ hartrees and the IP $=0.977$ hartree.
The soft-core potential with $Z=1$ approximates the situation where one electron
locates at the origin and shields the nuclear charge. 
The ePES1 has the potential depth close to that of the soft-core potential with $Z=1$
but with a broader potential well.
The ePES2 is more than twice as deep.
The ePES for He$^{+}$, given by Eq. (\ref{eq:ePEShatom}) with $Z=2$, is also included in the figure
for comparison.
The potential depth for ePES2 is deeper than that of the soft-core potential with $Z=2$
and is close to that of the ePES for He$^{+}$, 
but the asymptotic behavior in the long distance is closer to the potentials with $Z=1$.

Figure \ref{fig:he1dV} (b) displays the potential energy curves under the maximum laser field.
The height and width of the energy barrier of ePES1 are about 0.8 hartree and 10 bohrs.
Those for ePES2 are much higher and broader; about 4 hartrees and 30 bohrs.
We thus consider the ePES1 describes the active electron 
in response to the strong laser field.
Recall that the symmetry of the electronic wave function is taken into account as Eq. (\ref{eq:HL})
in the calculation to construct the ePES.

\subsubsection{High-harmonic generation spectra}

\begin{figure}[h]
\centering
\includegraphics[width=0.35\textwidth]{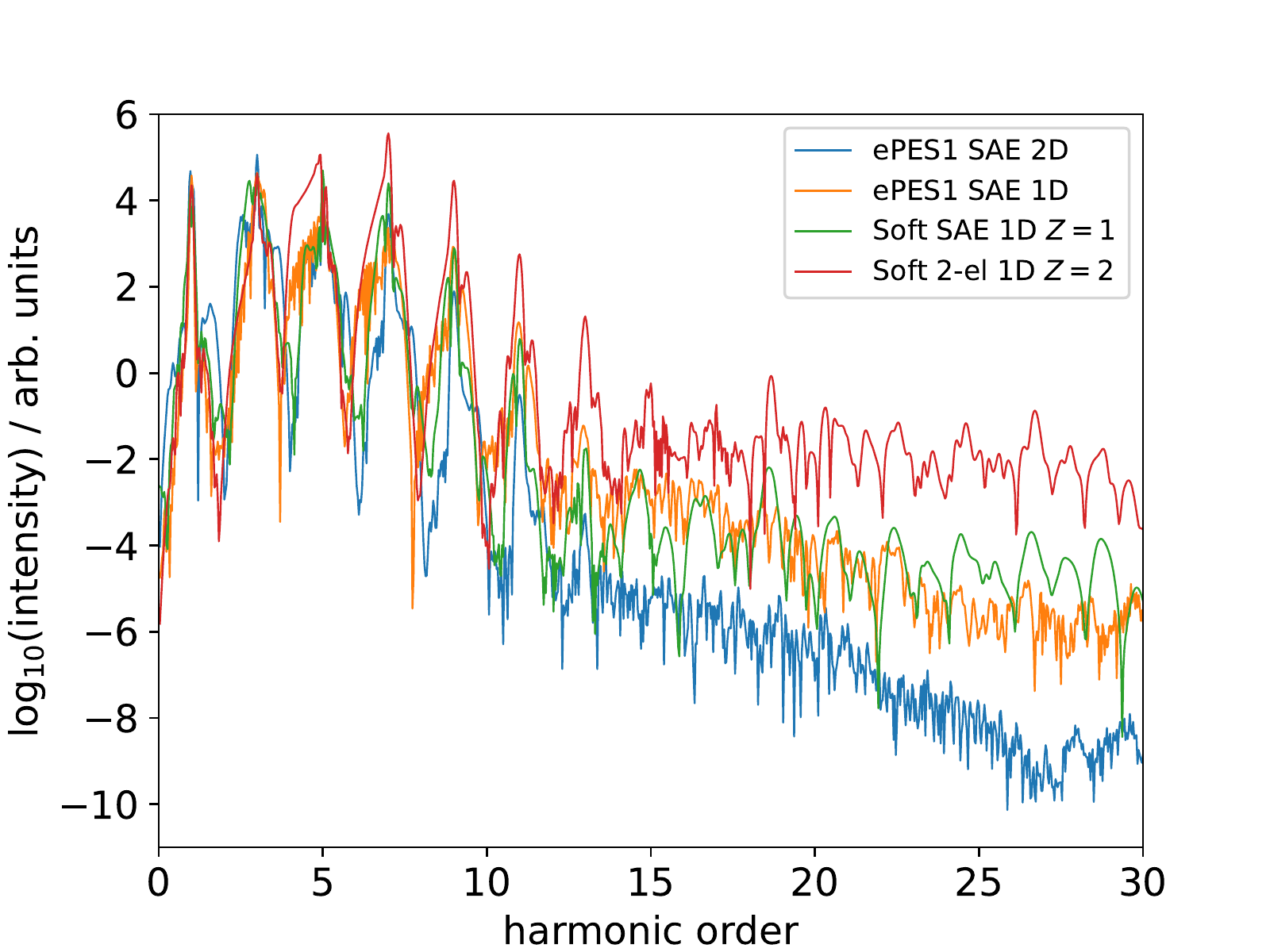}
\caption{High-harmonic generation spectra from helium atom
    with the single active electron (SAE) approximation on one- and two-dimensional ePES1,
    the SAE on one-dimensional soft-core potential Eq. (\ref{eq:vsoft}) 
    with $\epsilon=0.55$ and the nuclear charge $Z=1$,
    and the one-dimensional two electron calculation on the soft-core potential with
    $\epsilon=0.55$ and $Z=2$.}
\label{fig:heHHG}
\end{figure}

For the calculation of the HHG spectra,
the parameters of laser pulse in Eq. (\ref{eq:laserfield}) were taken from Ref. \cite{Guan2006}:
$\omega_0$ corresponding to the wavelength of 248.6 nm, 
the intensity $I_0 = 5 \times 10^{14} \;\mathrm{W/cm^2}$ 
giving ${\cal E}_0 = 6.138 \times 10^{8} \;\mathrm{V/cm}$
($= 1.194 \times 10^{-1}$ atomic units),
and $\tau = 24.88 \;\mathrm{fs}$ corresponding to 30 cycles.

Figure \ref{fig:heHHG} displays the computed HHG spectra.
They qualitatively agree with those in Ref. \cite{Guan2006};
clear peaks at odd harmonic orders and lack of apparent plateau and cut-off.
The latter features are explained by the short wave length of the input laser field
which gives
$U_p = 0.106$ hartree
and $\mathrm{IP} + 3.2 U_p = 6.8$ harmonic orders.
The two-dimensional calculation exhibits smallest intensity in the region higher than 11 harmonic orders.
This will be analyzed
with respect to the wave functions (probability densities) in Figs. \ref{fig:heWfn} and \ref{fig:he2dWfn}.
On the other hand, the one-dimensional two electron calculation with the soft-core potential,
the same model as in Ref. \cite{Bauer1997},
exhibits largest intensity in the high energy region.
The effects of the spatial dimension and the electron correlation thus appear to compensate each other.
As a result, the spectra from the one-dimensional SAE model with ePES1 and soft-core potential
have intermediate intensities in the high energy region and are close to each other.
The spectra in the low energy region up to 11 harmonic orders appear insensitive to the model.

\subsubsection{Tunnel probability density}

\begin{figure}[h]
\centering
\includegraphics[width=0.35\textwidth]{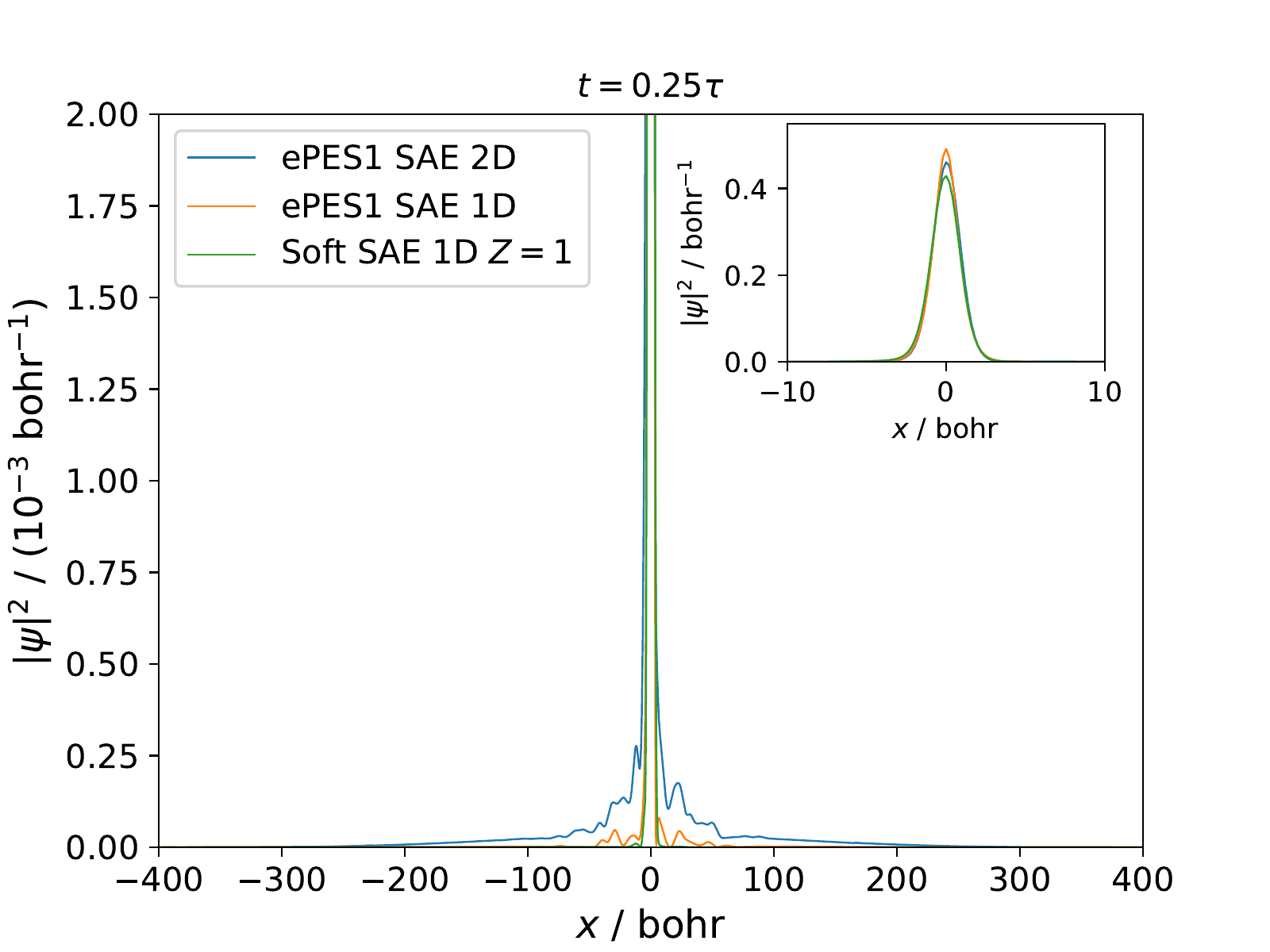}
\includegraphics[width=0.35\textwidth]{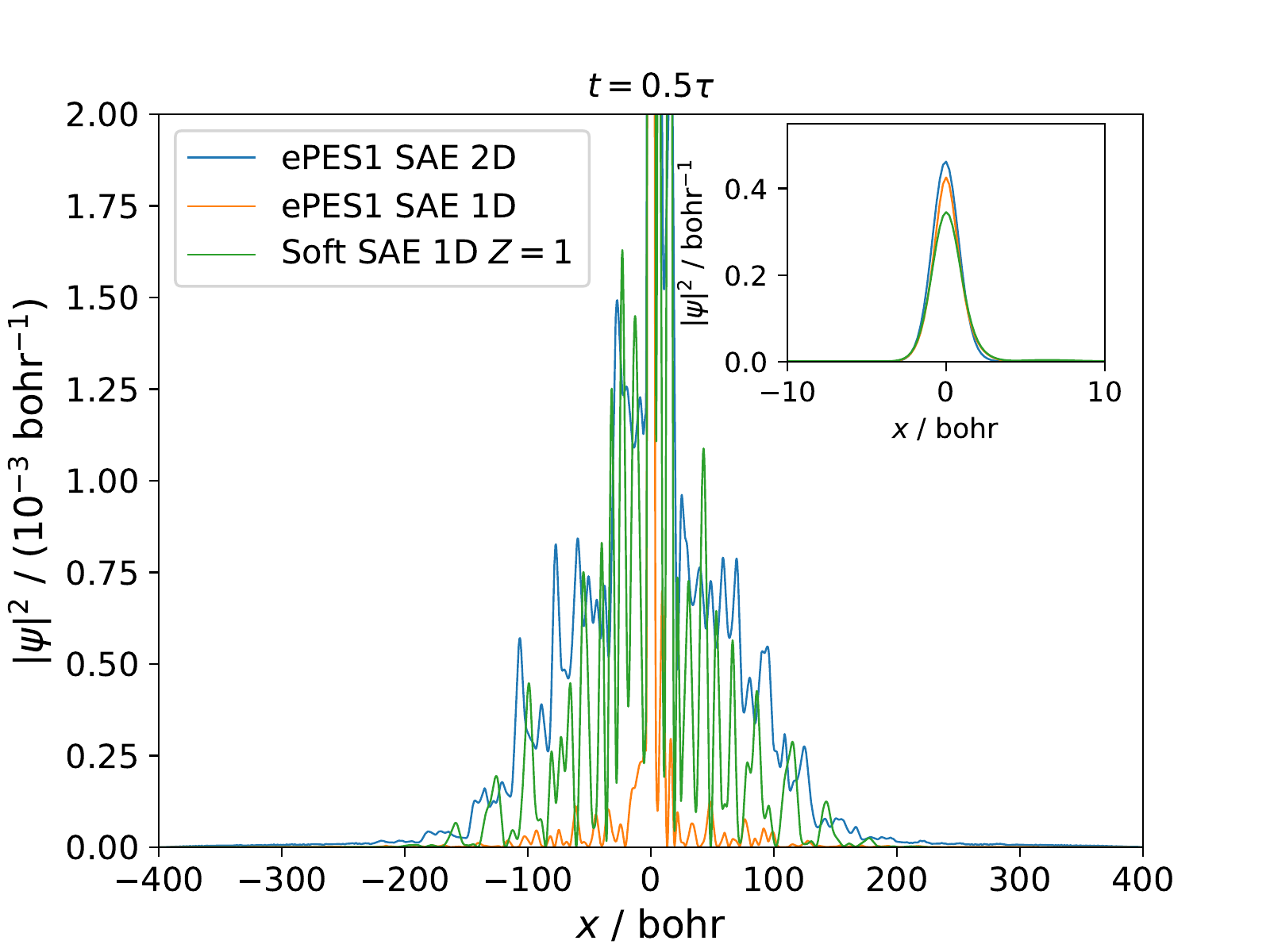}
\includegraphics[width=0.35\textwidth]{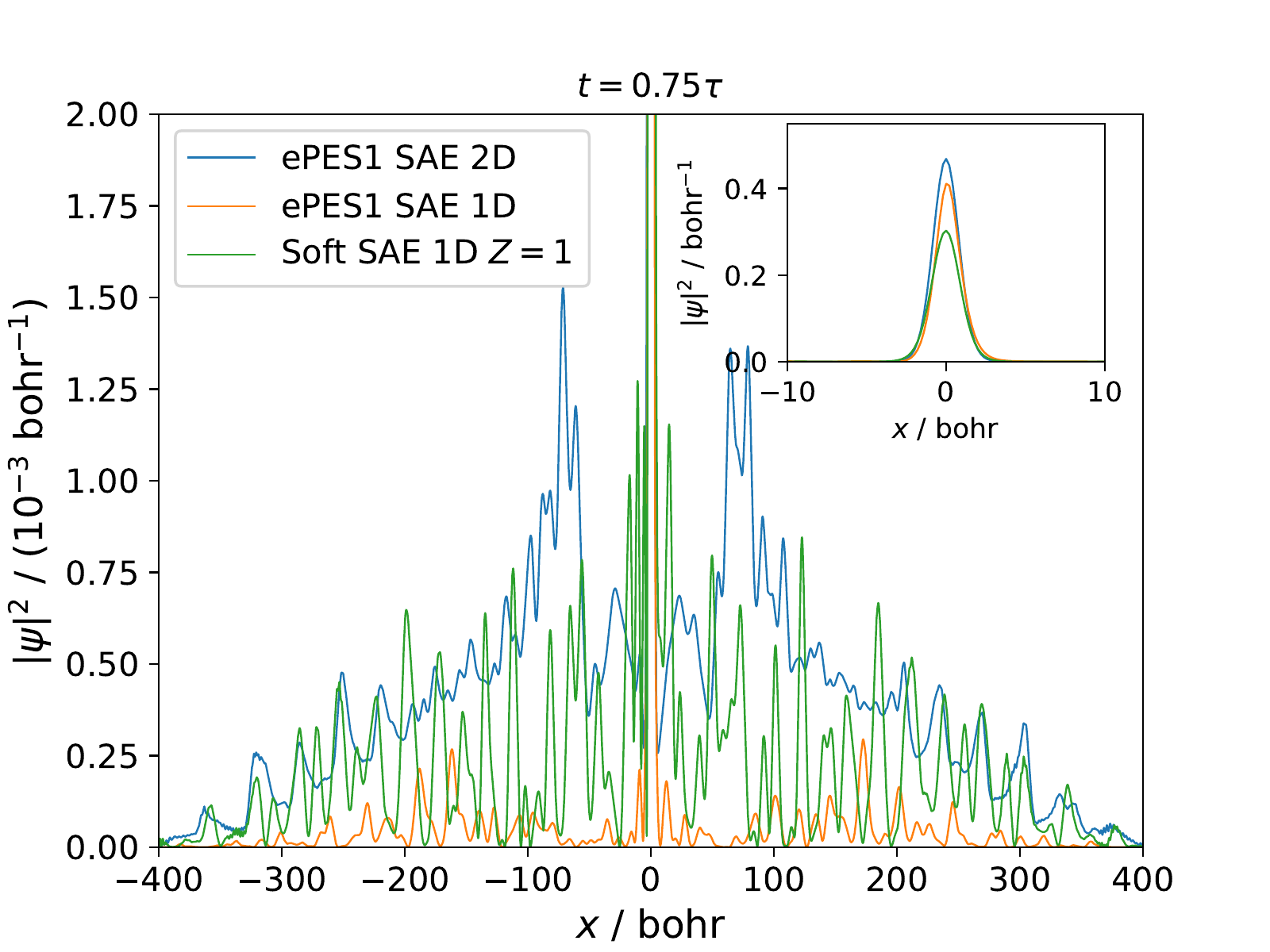}
\caption{Electron probability density of helium atom at $t=0.25\tau$, $0.5\tau$, and $0.75\tau$
with the single active electron (SAE) approximation on the one- and two-dimensional ePES1 
and the soft-core potential Eq. (\ref{eq:vsoft}) with $\epsilon=0.55$ and the nuclear charge $Z=1$.
The probability density of two-dimensional ePES1 calculation was integrated along the $y$ direction.}
\label{fig:heWfn}
\end{figure}

Figure \ref{fig:heWfn} displays the electron probability density at $t=0.25\tau$, $0.5\tau$, and $0.75\tau$.
The basic picture is similar to that for hydrogen atom discussed with Fig. \ref{fig:hWfn};
the electron probability density, which is orders of magnitude smaller 
than the central main distribution,
escapes through the potential barrier (Fig. \ref{fig:he1dV} (b)) by laser-induced quantum tunneling 
and spreads over several hundred bohrs.
The tunnel density for the two-dimensional ePES1 is larger than that
for the one-dimensional ePES1.

\begin{figure}[h]
\centering
\includegraphics[width=0.35\textwidth]{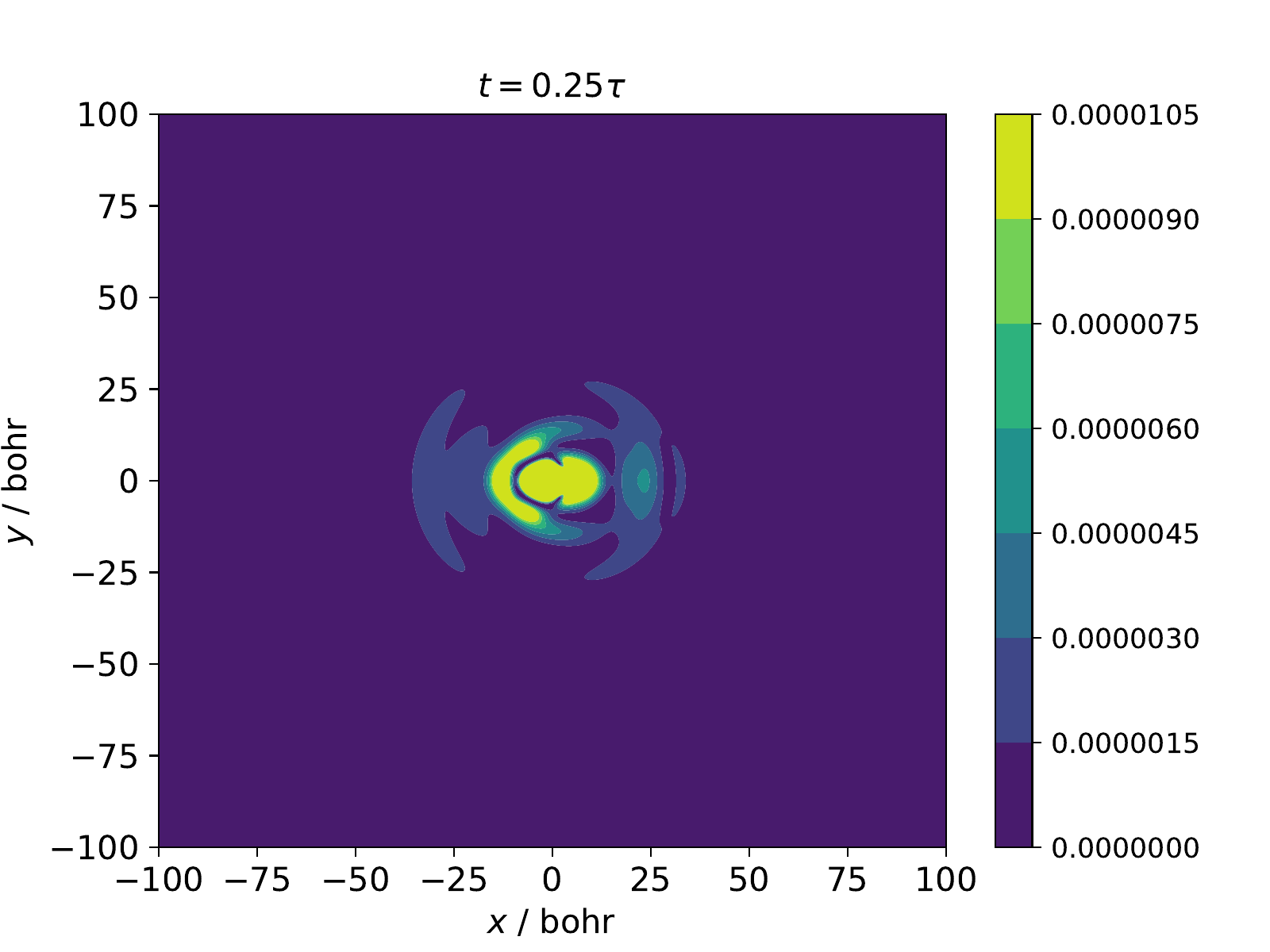}
\includegraphics[width=0.35\textwidth]{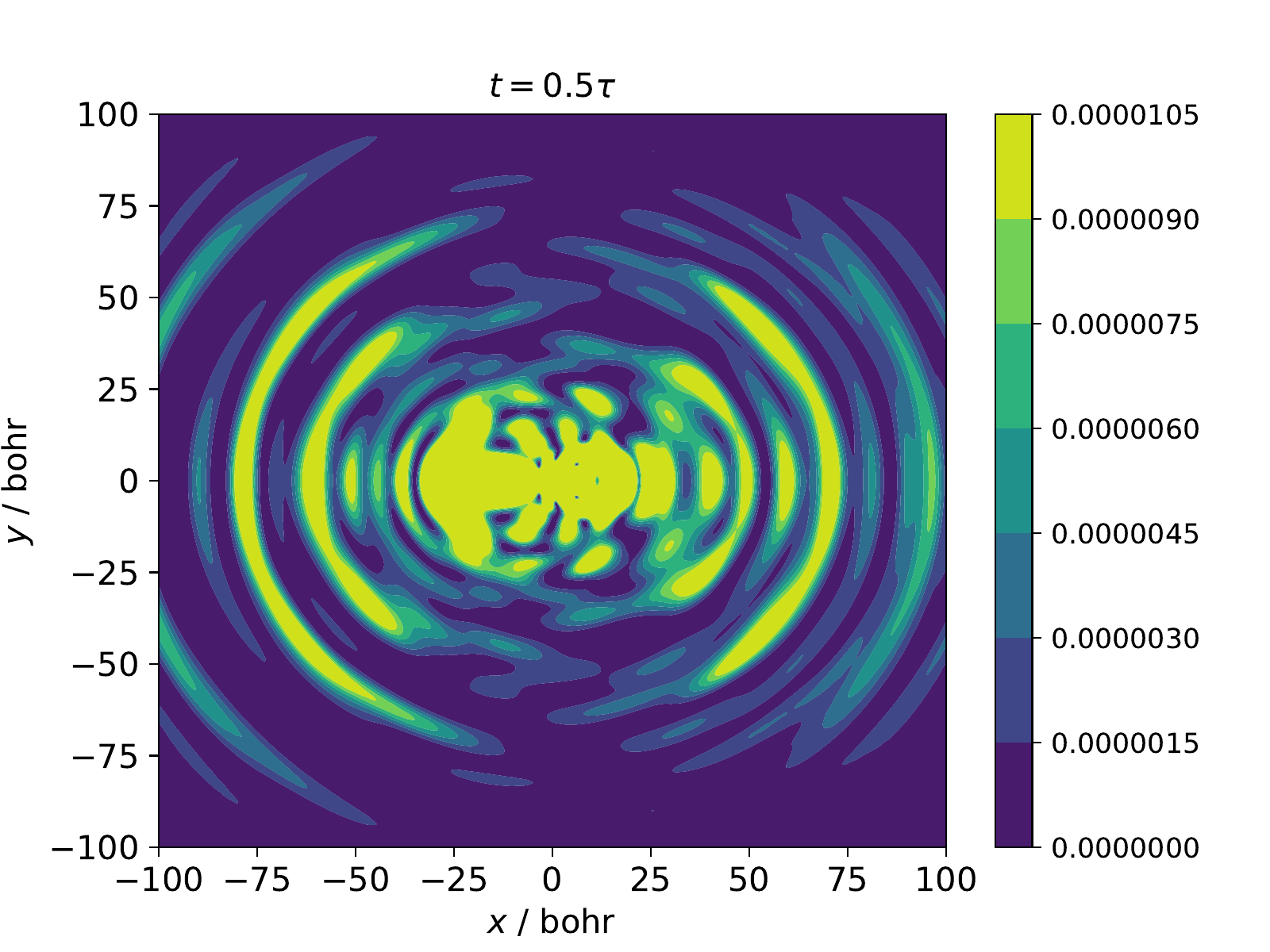}
\includegraphics[width=0.35\textwidth]{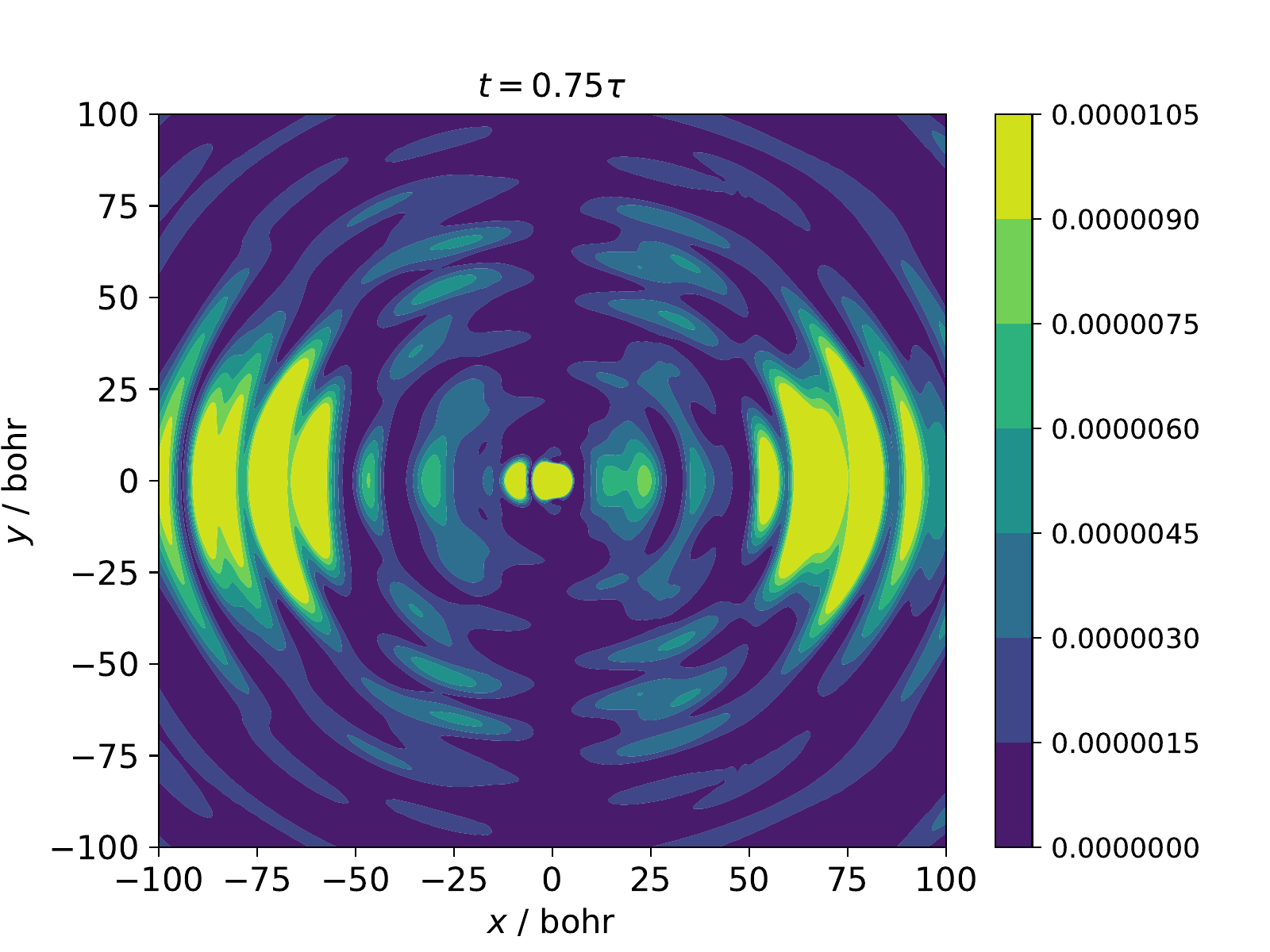}
\caption{Electron probability density of helium atom at $t=0.25\tau$, $0.5\tau$, and $0.75\tau$,
with the single active electron (SAE) approximation on the two-dimensional ePES1,
truncated at a maximum value set to $1\times 10^{-5}$ bohr$^{-2}$
to see the small component escaped from the potential well by the laser-induced quantum tunneling.
}
\label{fig:he2dWfn}
\end{figure}

Figure \ref{fig:he2dWfn} displays the electron probability density 
from the two-dimensional SAE calculation on the ePES1 at $t=0.25\tau$, $0.5\tau$, and $0.75\tau$.
They are truncated at a maximum value set to $1\times 10^{-5}$ bohr$^{-2}$
to see the small component outside the potential well.
The basic picture is again similar to that for hydrogen atom discussed with Fig. \ref{fig:h2dWfn};
the component distributed toward $y$-direction
appears to suppress the HHG higher than 11 harmonic orders in Fig. \ref{fig:heHHG}
compared to the one-dimensional calculations.

\subsubsection{Dipole dynamics and spectral evolution}

\begin{figure}[h]
\centering
\includegraphics[width=0.35\textwidth]{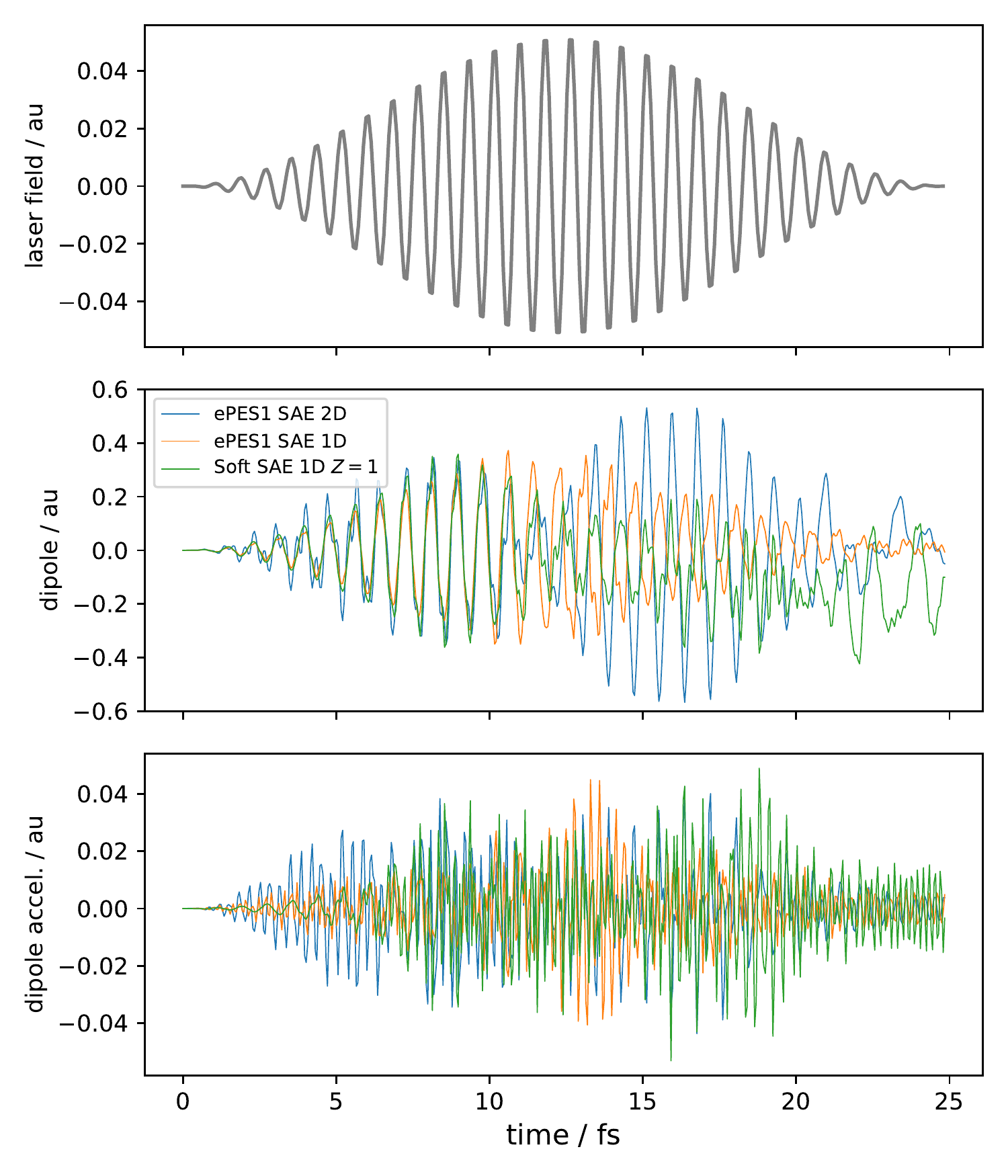}
\caption{Time profiles of the laser field (top) and the dipole moment (middle)
and the dipole acceleration (bottom) of helium atom,
with the single active electron (SAE) approximation on the one- and two-dimensional ePES1 
and the soft-core potential Eq. (\ref{eq:vsoft}) with $\epsilon=0.55$ and the nuclear charge $Z=1$.}
\label{fig:heDplt}
\end{figure}

Figure \ref{fig:heDplt} displays the time profiles of the laser pulse, dipole moment, and dipole acceleration.
The dipole dynamics of the different models are similar up to $t \simeq 10$ fs.
The amplitude of the dipole oscillation for the two-dimensional ePES1 model
is notably larger after $t \simeq 13$ fs. 
Then, the phase of the dipole oscillation of the one-dimensional ePES1 model shift almost by $\pi$
compared to the other two models.
The dipole acceleration dynamics of the different calculations are also notably different.
These are, however, not obviously reflected in the HHG spectra in Fig. \ref{fig:heHHG}.

\begin{figure}[h]
\centering
\includegraphics[width=0.35\textwidth]{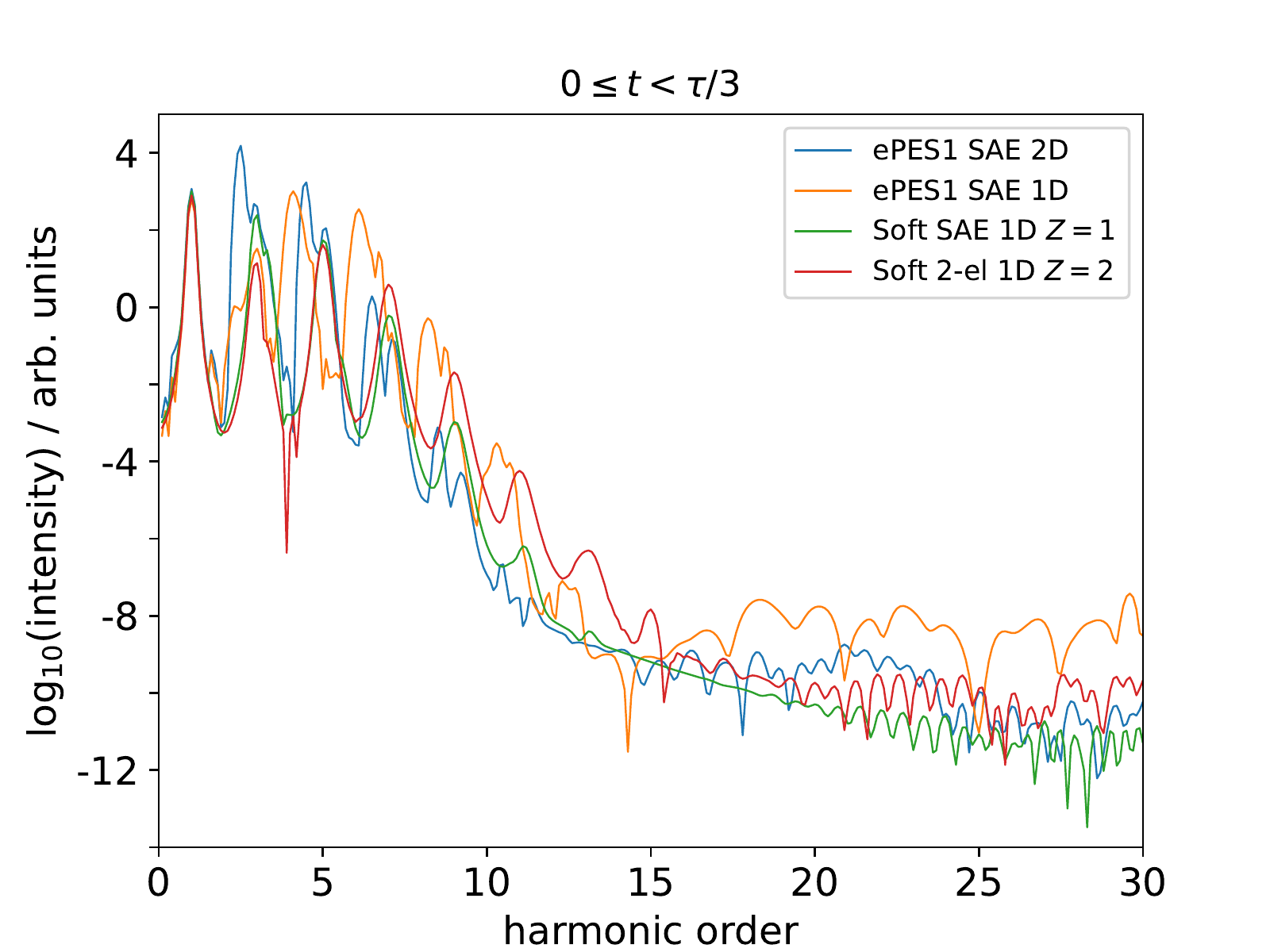}
\includegraphics[width=0.35\textwidth]{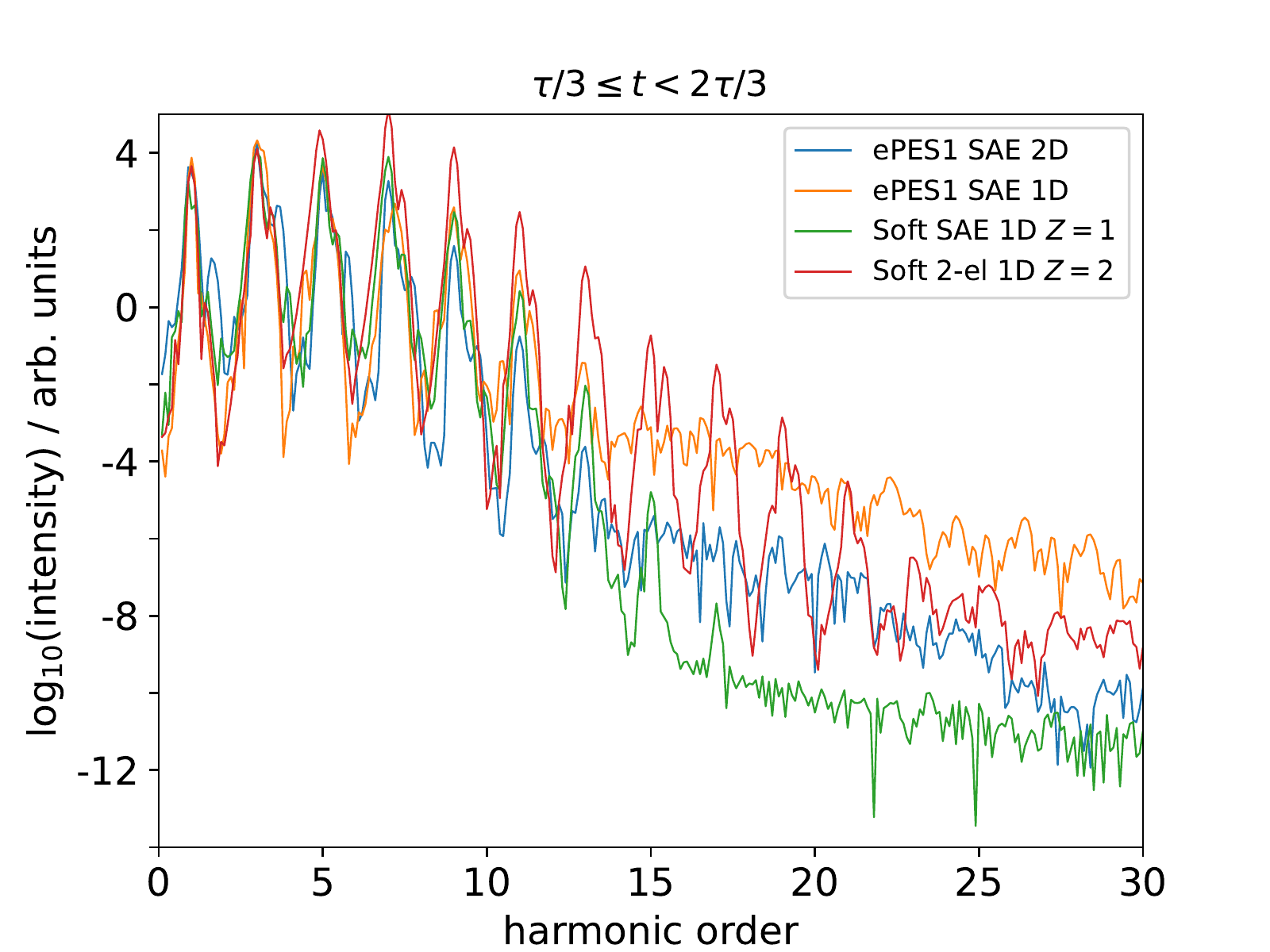}
\includegraphics[width=0.35\textwidth]{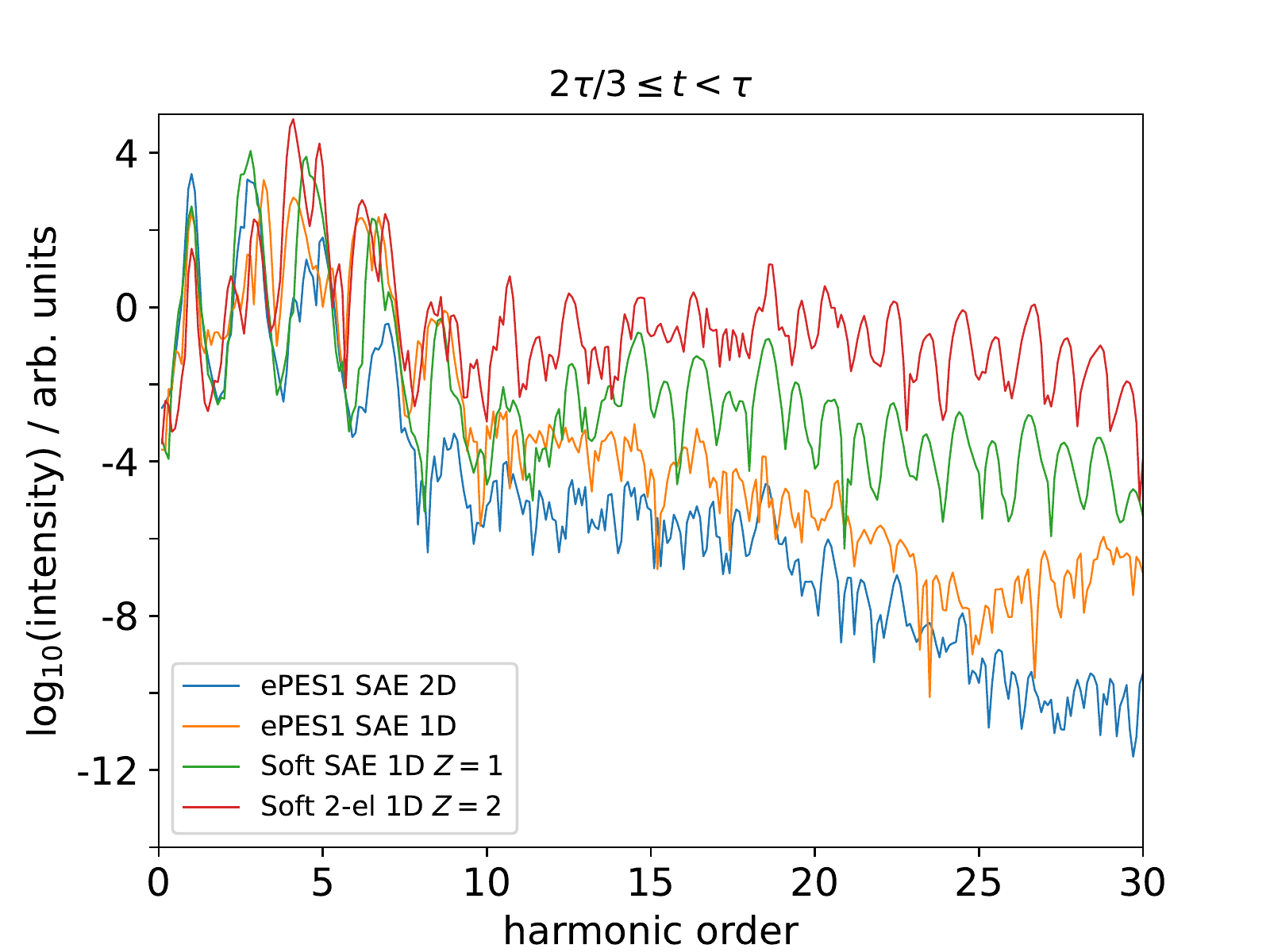}
\caption{High-harmonic generation spectra of helium atom,
similar to Fig. \ref{fig:heHHG} but 
computed from three time windows, 
$0 \le t < \tau/3$, $\tau/3 \le t < 2\tau/3$, and $2\tau/3 \le t < \tau$.}
\label{fig:heHHG3tw}
\end{figure}

The HHG spectra computed from the three time windows,
$0 \le t < \tau/3$, $\tau/3 \le t < 2\tau/3$, and $2\tau/3 \le t < \tau$,
are displayed in Fig. \ref{fig:heHHG3tw}.
In $0 \le t < \tau/3$,
the spectra of the one- and two-dimensional ePES1 models show peaks 
at even harmonic orders,
but the overall profiles of the spectra are comparable among the models.
In $\tau/3 \le t < 2\tau/3$,
the main peaks up to 11 harmonic orders are similar among the models as in Fig. \ref{fig:heHHG}.
Notably, the two-electron model with the soft-core potential 
shows clear peaks up to 21 harmonic orders.
This seems to come from coherent motion of correlated two electrons under the high laser intensity.
The spectrum for the soft-core potential shows clear peaks up to 17 harmonic orders,
but the intensity is the smallest in the high harmonic region.
In $2\tau/3 \le t < \tau$,
the main peaks in the low harmonic region are rather disordered with even and non-integer harmonic orders.
The behavior in the high harmonic region is qualitatively similar to Fig. \ref{fig:heHHG}.
Interpretation of these results in relation to the potential shapes in Fig. \ref{fig:he1dV}
does not seem trivial.

\section{Concluding Remarks}
\label{sec:concl}
Many previous studies demonstrated the usefulness of SAE approximation 
for calculation of electron dynamics and HHG spectra under intense laser fields.
In particular, the soft-core potential conveniently provides a SAE model with single parameter, 
although the model seems rather ad-hoc and 
the parameter determination involves arbitrariness.
There are attempts to construct SAE models 
by reducing accurate many electron wave functions and energies
to effective one electron models
\cite{Ohmura2017,Kato2018,Kocak2020,Schild2017}.
The present model of localized eWP with VB coupling offers a simple and intuitive method 
to construct an effective one electron potential (ePES) 
without adjustable parameters.
Interestingly, in the case of helium atom, 
the VB wave function consists of two eWP with different sizes and binding energies
while maintaining the overall symmetry.
The eWP with smaller binding energy describes the active electron responding to the strong laser field.
This work confirmed the usefulness of the ePES model for hydrogen and helium atoms.
The basic picture that emerged
on the mechanism of HHG would extend the conventional semi-classical three-step model
to the regime where the laser-induced quantum tunneling dominates:
the main part of the wave function is bound near the nuclei
while the parts orders of magnitude smaller in probability density 
that escaped through the potential barrier by laser-induced
quantum tunneling mainly contribute to the HHG.

One of our original aims of studying the ePES was to construct 
a model of electronic polarization in condensed phase molecular dynamics simulations
combined with nuclear wave packets for light atoms
\cite{Kim2014,Kim2014prb,Kim2015,Kim2016,Abe2018,Yamaoka2021}.
Work along this line is under way.

\section*{Acknowledgments}
This work has been supported by JSPS KAKENHI No. 19K22173.

\end{document}